\documentclass{egblank}
\usepackage{pg2025}
\usepackage[T1]{fontenc}
\usepackage{dfadobe}  

\usepackage{cite}  %
\newcommand{\citet}[1]{\cite{#1}}
\newcommand{\citep}[1]{\cite{#1}}

\newcommand{\citeauthor}[1]{\cite{#1}}
\newcommand{\citeyear}[1]{\cite{#1}}

\BibtexOrBiblatex

\electronicVersion
\PrintedOrElectronic

\usepackage{egweblnk} 

\usepackage{graphicx}
\usepackage{amsmath,amssymb}
\usepackage{caption}
\usepackage{subcaption}
\usepackage{xcolor}
\usepackage{lipsum}
\usepackage{tikz}      %
\usepackage{tabularx}
\usepackage{booktabs}
\usepackage{array}
\usepackage[ruled,vlined]{algorithm2e}
\usepackage{pbox}

\definecolor{darkgreen}{rgb}{0.0, 0.5, 0.0}

\definecolor{wip_orange}{rgb}{1.0, 0.5, 0.0} %

\newif\ifrevisions
\revisionsfalse

\newenvironment{revision}{\begingroup\ifrevisions\color{red}\fi}{\endgroup}

\title{Pro-DG: Procedural Diffusion Guidance for Architectural Facade Generation}

\author[A. Plocharski \& J. Swidzinski \& P. Musialski]
{\parbox{\textwidth}{\centering
    A. Plocharski\orcid{0000-0002-7487-8153}\thanks{aleksander.plocharski@pw.edu.pl}$^{1,2}$,
    J. Swidzinski\orcid{0009-0001-0511-2058}\thanks{janswidzinski01@gmail.com}$^{3}$ and
    P. Musialski\orcid{0000-0001-6429-8190}\thanks{przem@njit.edu}$^{4}$
        }
        \\
{\parbox{\textwidth}{\centering
$^1$Warsaw University of Technology, Poland\\
$^2$Akces NCBR, Poland\\
$^3$Imperial College London, England\\
$^4$New Jersey Institute of Technology, United States
       }
}
}

\begin{document}

\teaser{
    \centering
    \begin{minipage}[c]{0.33\textwidth}
        \centering
        \includegraphics[height=4.4cm]{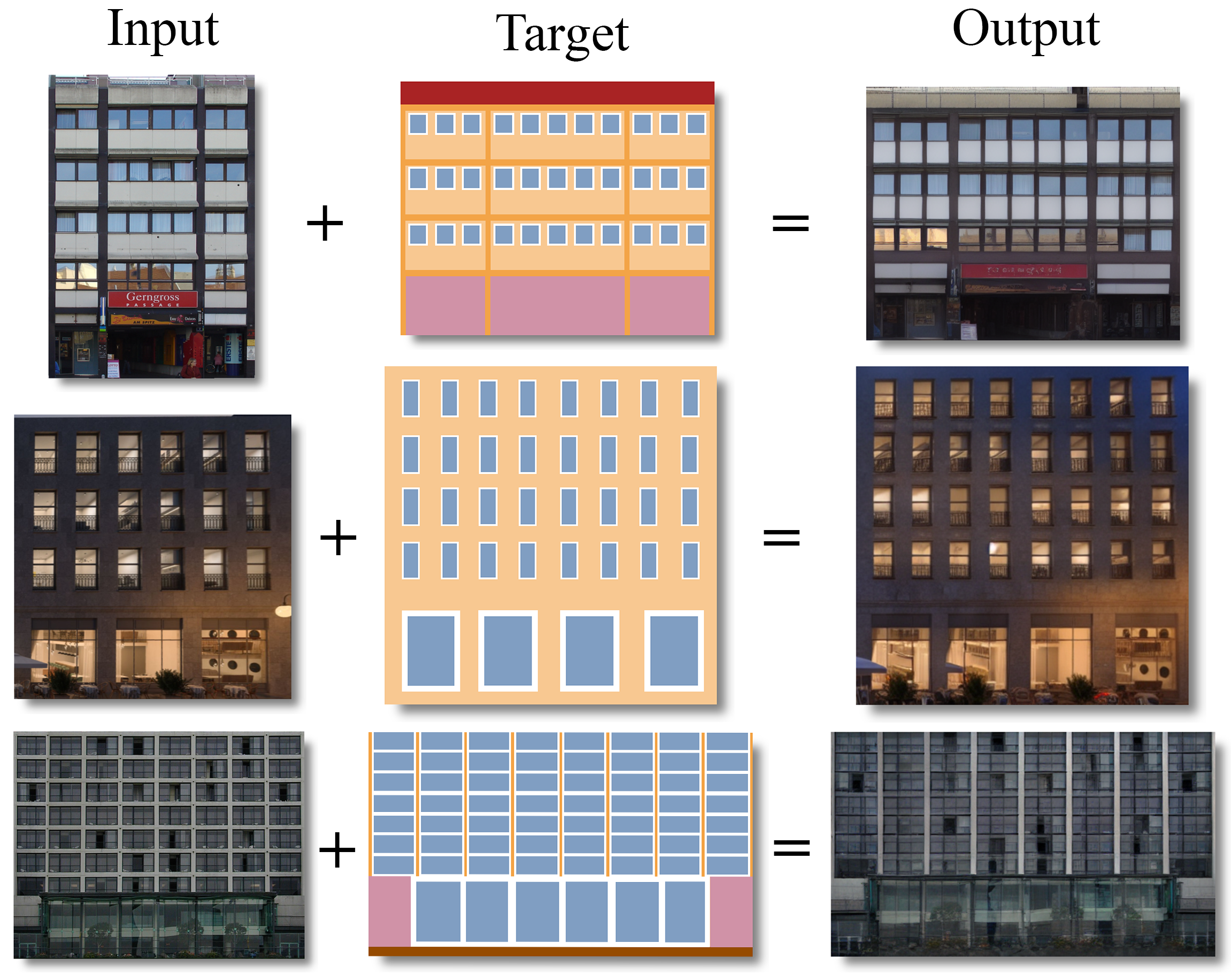}
    \end{minipage}%
    \hspace{0.007\textwidth} %
    \begin{minipage}[c]{0.005\textwidth}
        \centering
        \rule{0.5pt}{4cm} %
    \end{minipage}%
    \hspace{0.007\textwidth} %
    \begin{minipage}[c]{0.63\textwidth}
        \centering
        \raisebox{0.0cm}{\includegraphics[height=3.7cm]{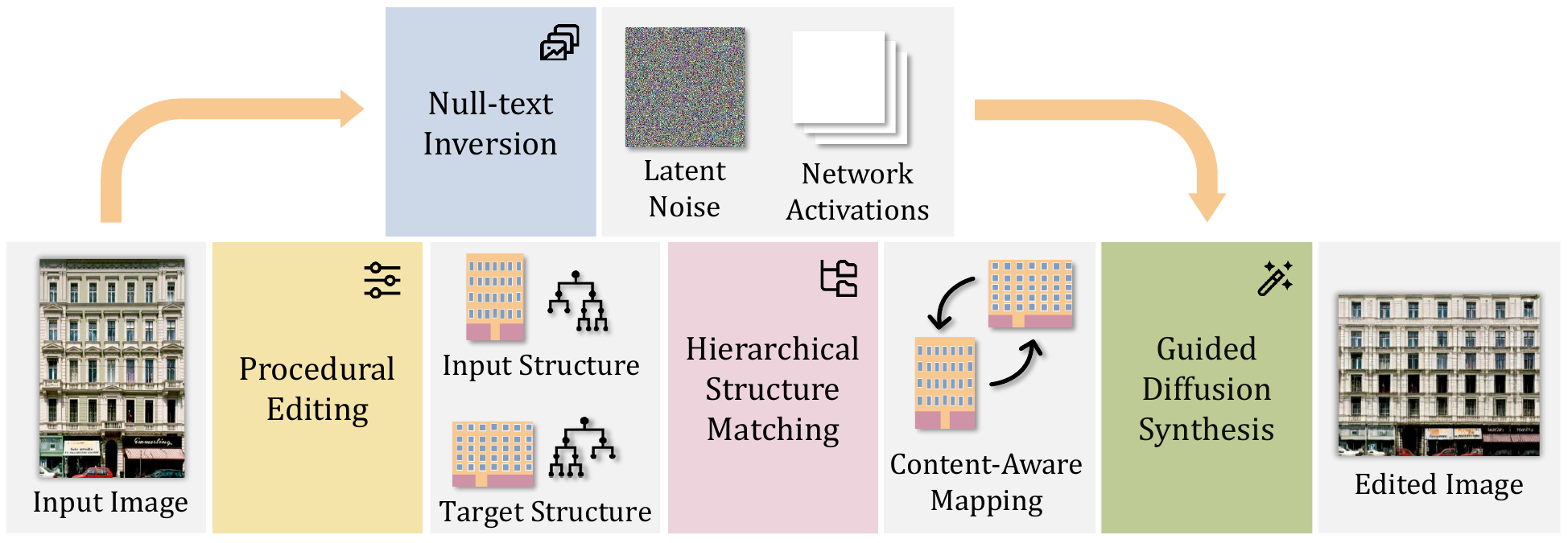}}
    \end{minipage}
    \caption{Given a real facade image and a target grammar-based procedural structure, our method generates layout-consistent realistic edits that preserve the core identity of the original design (left). Our pipeline combines symbolic procedural modeling with diffusion synthesis for controllable, structure-aware image editing (right).}
    \label{fig:teaser}
}

\maketitle
\begin{abstract}
We use hierarchical procedural rules for the generation of control maps within the stable diffusion framework to produce photo-realistic architectural facade images. Starting from a single input image and its segmentation, we apply an inverse procedural module to identify the facade's hierarchical layout. Leveraging this hierarchy and structural features, we introduce a novel ControlNet pipeline that generates new facade imagery guided by procedural transformations. Our method enables various structural edits, including floor duplication and window rearrangement, by integrating hierarchical alignment directly into control maps. This precisely guides the diffusion-based generative process, ensuring local appearance fidelity alongside extensive structural modifications. Comprehensive evaluations, including comparisons with inpainting-based approaches and synthetic benchmarks, confirm our approach’s superior capability in preserving architectural identity and achieving accurate, controllable edits. Quantitative results and user feedback validate our method's effectiveness.


\end{abstract}

\section{Introduction}
\label{sec:introduction}
Facade design intricately balances aesthetics, functionality, and structural coherence, serving as a vital component of architectural heritage and modern urban landscapes alike. Automating the generation of realistic facades that adhere to architectural principles while offering user-driven flexibility remains a significant challenge. What proves even more challenging is generating coherent variations of already existing facade designs.

Previous methods in graphics and vision focused either on procedural modeling~\cite{wonka2003instant} or facade-parsing methods~\cite{teboul2013shape}, while recent work introduced neuro-symbolic reconstruction~\cite{facaid2024}. 
At the same time, recent advancements in image synthesis through \emph{denoising diffusion models} \citep{ho2020denoising, song2020denoising} have made significant advances in photo-realistic asset generation. One still challenging task is the control over the resulting outputs of such models.

Such control is especially of interest for well-structured images, like architectural models, where precision and accuracy are of high importance. 
Specifically, maintaining \emph{structural consistency} to ensure that facade elements such as windows, doors, and balconies are coherently arranged; providing \emph{controllability} that allows users to manipulate specific design elements; and achieving \emph{photorealism} that preserves realistic textures, lighting, and materials, all while adhering to user-defined procedural rules and real-life reference designs.

We address the largely unexplored gap between symbolic procedural structures and diffusion models with a framework that fuses procedural facade modeling with diffusion-based image synthesis. Starting from a single input image and its segmentation, we reconstruct its procedural structure, apply user-driven edits, and guide a diffusion model to produce realistic, structurally consistent edits that retain the core visual identity of the original (Figure~\ref{fig:teaser}). Our method showcases the untapped potential of combining symbolic, hierarchical structures with diffusion models for high-quality, controllable image generation. Additionally, contrary to most state-of-the-art diffusion-based editing methods, we focus on restructuring the whole image-space instead of transforming only specific objects in the image.

\textbf{Contributions.} We present a unified framework that couples symbolic facade grammars with diffusion models for \emph{a holistic, structured, and controlled} facade-image editing. Our contributions are:
(i) a hierarchy-aware matching algorithm that enforces parent–child consistency while combining structural and visual cues; 
(ii) a pipeline that synthesizes hierarchy-derived control signals and activation targets to guide all diffusion steps; and
(iii) an evaluation across diverse facade-images demonstrating consistent gains in structural fidelity and visual coherence.\\
When constructing our unified pipeline, we utilize Fa\c{c}AID~\cite{facaid2024} and Null-text Inversion~\cite{nulltext2023} as given building blocks, and use Diffusion Handles~\cite{pandey2024diffusionhandles} as the starting point for our guided-inference algorithmic module. In our guidance system, we (a) expand their underlying network and its conditioning to enable hierarchy-derived control, (b) augment their correspondence estimation with our hierarchy-based mapping, and (c) redesign the optimization objective and their step-wise schedule. Figure~\ref{fig:pipeline} presents our unified pipeline: modules with \textbf{bold borders} are entirely novel components introduced in his work, and {dashed borders} denote novel components {adapted from prior work} (Guided Inference, based on Diffusion Handles).

The goal is to streamline the workflow for facade design: from high-level specification and manipulation of structural elements to automatic photorealistic editing. In doing so, we unite the expressiveness of procedural grammars with the flexibility and generative prowess of diffusion models, enabling a new level of user-driven creativity and control in facade synthesis.

The remainder of this paper is organized as follows. Section~\ref{sec:related_work} surveys related works in procedural modeling and diffusion-based image synthesis and editing. Section~\ref{sec:method} describes the details of our pipeline, how hierarchical structures are translated into image space mappings and used for guiding the diffusion process. Section~\ref{sec:evaluation} presents our experimental setup and results, highlighting the benefits of our method. Finally,  Section~\ref{sec:conclustions} provides a discussion of limitations, and potential extensions of this research.

\section{Related Work}
\label{sec:related_work}

\paragraph{Facade Modeling.}
Procedural facade modeling began with shape grammars \citep{stiny1975shape}, enabling both large-scale synthetic cities \citep{parish2001procedural, wonka2003instant, mueller2006procedural} and \emph{inverse procedural modeling} to infer grammar rules from real data \citep{aliaga2007style, ripperda2009application, stava2010inverse, teboul2011automated, musialski2013survey}. \begin{revision}Other classic graphics approaches tackled this by employing by-example synthesis for architectural textures \citep{Lefebvre2010} or optimizing structure completion for facade layouts \citep{Fan2014}\end{revision}. Despite semi-automatic reconstruction, achieving photorealistic facades typically required substantial artist intervention. Recent advances combine deep learning with procedural methods: \citet{mathias2011facade} trained facade priors for grammar-driven splits, \citet{teboul2013shape} refined symbolic expansions via machine learning, and \citet{facaid2024} introduced a neuro-symbolic pipeline for learned facade grammars. Still, reconciling rule-based formalisms with contemporary neural generators remains challenging. \begin{revision}Notably, the foundational idea of combining procedural generative models with synthesis techniques to enforce structural consistency was earlier explored in the domain of texture synthesis \citep{Guehl2020}. While these prior work successfully utilized by-example and semi-procedural textures, our framework expands on this philosophy by uniquely integrating symbolic hierarchies directly into the latent space of modern diffusion models, ensuring global structural coherence during large-scale architectural edits.\end{revision}%

\paragraph{Diffusion Models for Image Synthesis.}
Recent advances in diffusion models \cite{sohl2015deep,ho2020denoising,song2020denoising} have revolutionized image synthesis by iteratively denoising noisy inputs to generate high-fidelity images. Latent diffusion frameworks \cite{rombach2022highresolution} further improve computational efficiency while preserving image quality, and text-conditioned variants \cite{saharia2022photorealistic,ramesh2022hierarchical} enable detailed semantic control for text-to-image generation. Methods such as ControlNet \cite{zhang2023adding} augment the process by conditioning on auxiliary inputs like edge maps or segmentation masks, though scaling these approaches to complex and repetitive domains—such as multi-story facades—remains challenging. More recent works extend these foundations by integrating additional conditioning signals: SpaText \cite{avrahami2023spatext} introduces a spatio‑textual representation that allows open‑vocabulary scene control, while ObjectStitch \cite{song2023objectstitch} demonstrates object compositing with diffusion models. Furthermore, MultiDiffusion \cite{bartal2023multidiffusion} proposes a unified framework that fuses multiple diffusion trajectories, enhancing user controllability over image synthesis without retraining. 

These developments collectively push the boundaries of diffusion-based synthesis and inform the structural guidance employed in our framework.

\paragraph{Controllable Image Editing.}
Early work in controllable image generation primarily focused on localized pixel-level edits or attribute manipulation via text prompts \cite{avrahami2022blended,pan2023drag,brooks2023instructpix2pix}. In contrast, more recent approaches have aimed for fine‑grained control over specific objects present in the images. For instance, Diffusion Handles \cite{pandey2024diffusionhandles} enables 3D‑aware edits by lifting intermediate activations to 3D space and applying rigid transformations, and methods like Zero‑1‑to‑3 \cite{liu2023zero123} enforce multi-view consistency through explicit geometric constraints. Additionally, diffusion based compositing methods like ObjectStich \cite{objectStitch}, can also be used for single object image edits.

Complementary to these, Training‑Free Layout Control with Cross‑Attention Guidance \cite{chen2024trainingfree} leverages cross‑attention maps to steer the generation process toward user‑specified layouts without additional training. T2I‑Adapter \cite{mou2023t2iadapter} \begin{revision}and IP-Adapter \cite{ye2023ip-adapter}\end{revision} further enhance control by aligning latent representations with external signals \begin{revision}or image prompts\end{revision}, while Grounding DINO \cite{liu2023groundingdino} provides robust open‑set object detection to accurately delineate regions of interest for subsequent procedural editing. Additionally, Object 3DIT \cite{3dit} integrates language‑driven 3D‑aware editing, offering a pathway for preserving object identity while modifying spatial configurations. 

Together, these advances underscore a growing trend toward integrating explicit, multi‑scale control mechanisms with diffusion models—a trend that our framework leverages by combining procedural shape grammars with diffusion-based synthesis to enforce both local appearance fidelity and global structural consistency. Additionally, in contrast to the most prevalent diffusion based image editing methods which focus moving a specific object in the image, our approach enables the edit to influence the whole image space while still staying true to the original image reference.

Our method employs \emph{procedural-level} control-leveraging a domain-specific grammar to define facade layouts with precision, rather than relying on purely pixel-based or language-based constraints. This neuro-symbolic fusion permits photorealistic editing while guaranteeing architectural integrity and addresses the longstanding gap between symbolic, rule-based modeling and data-driven diffusion methods, offering a powerful new framework for facade design, urban simulation, and beyond.

\section{Method}
\label{sec:method}

\paragraph{Problem Statement. }
The goal of our method is to enable procedurally modifying an image of a facade while preserving its core identity. Specifically, the edited image should remain recognizable as a variation of the original, rather than an entirely new design. The resulting image must also be realistic, architecturally plausible, and well-structured.

Let $F_{in}$ denote the input facade image, $E$ represent the user's edit, and $F_{out}$ be the output facade image after applying the edit. Our method models the function
$$\mathcal{M}(F_{in}, E) = F_{out},$$
which maps the input image and edit to a plausible output image.

\begin{figure}[t]
        \centering
        \includegraphics[width=0.98\columnwidth]{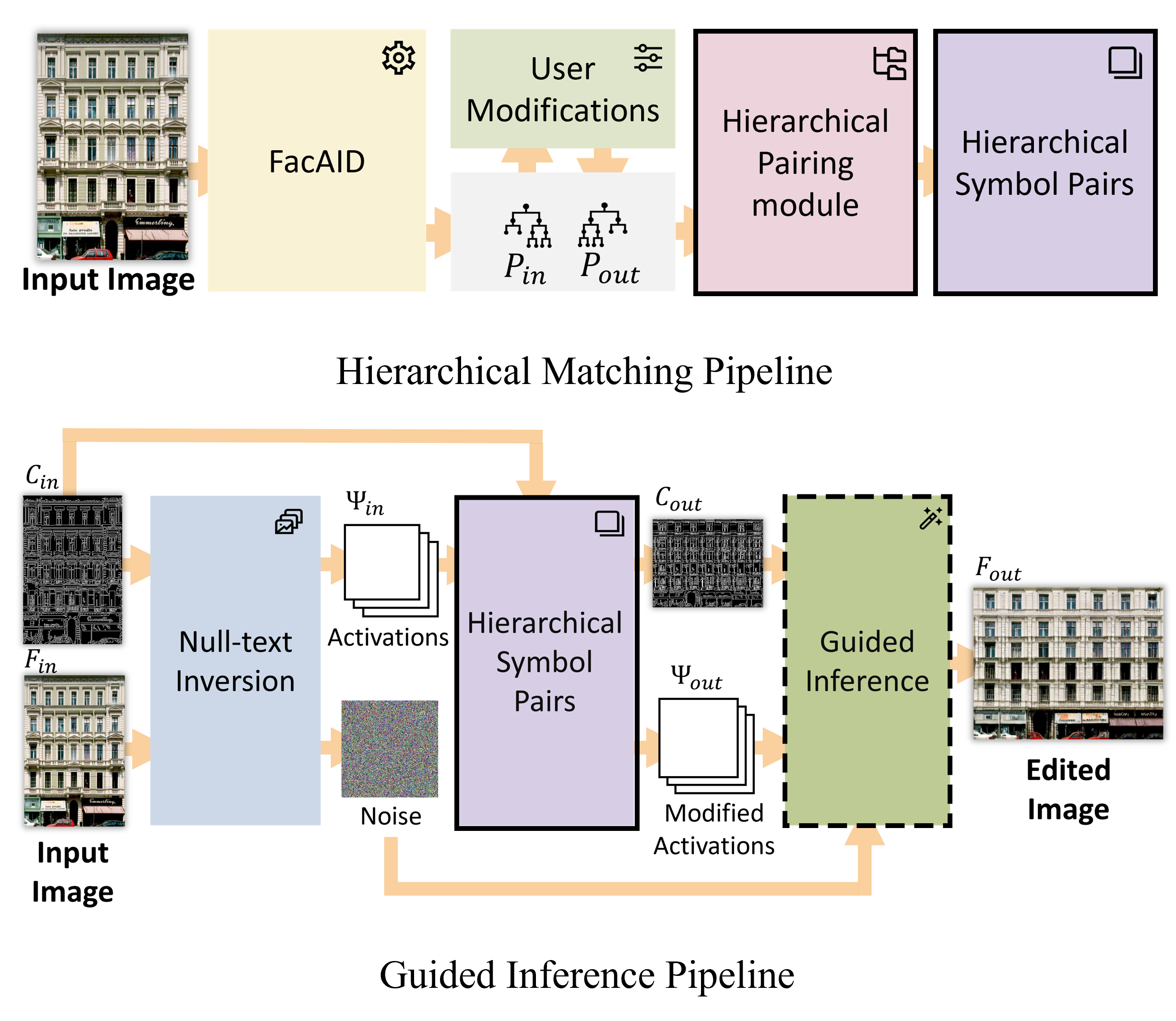}
        \caption{The pipeline consists of two distinct elements: the Hierarchical Matching Pipeline \& the Guided Inference Pipeline. The first one is responsible for finding the correspondences between the source and target image spaces using information encoded in the hierarchical representations while the second one guides the diffusion process based on those correspondences. Modules with bold borders are entirely novel components introduced in his work, and dashed borders denote novel components adapted from prior work.}
        \label{fig:pipeline}
\end{figure}

\paragraph{Method Overview. }
Our method consists of a back-to-back generation pipeline (cf. Figure~\ref{fig:pipeline}) with the following components:

\begin{enumerate}
    \item \textbf{Inverse Procedural Reconstruction.} Reconstructing the procedural facade representation from the input image $F_{in}$ and its segmentation.
    \item \textbf{Structure Editing.} Modifying the procedural representation in order to create the edit $E$ (user-driven).
    \item \textbf{Hierarchical Matching.} Creating a mapping between elements of the original facade structure and the one modified by edit $E$.
    \item \textbf{Diffusion Reconstruction.} Reconstructing the original image as a diffusion model output to get the noise which maps to the input image.
    \item \textbf{Edited Facade Inference.} Guiding the inference process of a diffusion model to generate $F_{out}$.
\end{enumerate}
Our system builds upon three methods from literature---Fa\c{c}AID \citep{facaid2024}, Null-text Inversion \citep{nulltext2023} and Diffusion Handles \cite{pandey2024diffusionhandles}---by combining their functionality into one coherent pipeline while also adding new crucial standalone elements and adapting those existing approaches to fit our use case. Additionally, we employ a Canny edges controlled ControlNet model \citep{zhang2023adding}---finetuned on 23k facade images from the LSAA dataset \citep{lsaa}---as a wrapper for our diffusion model.

\begin{figure}[t]
        \centering
        \includegraphics[width=0.95\columnwidth]{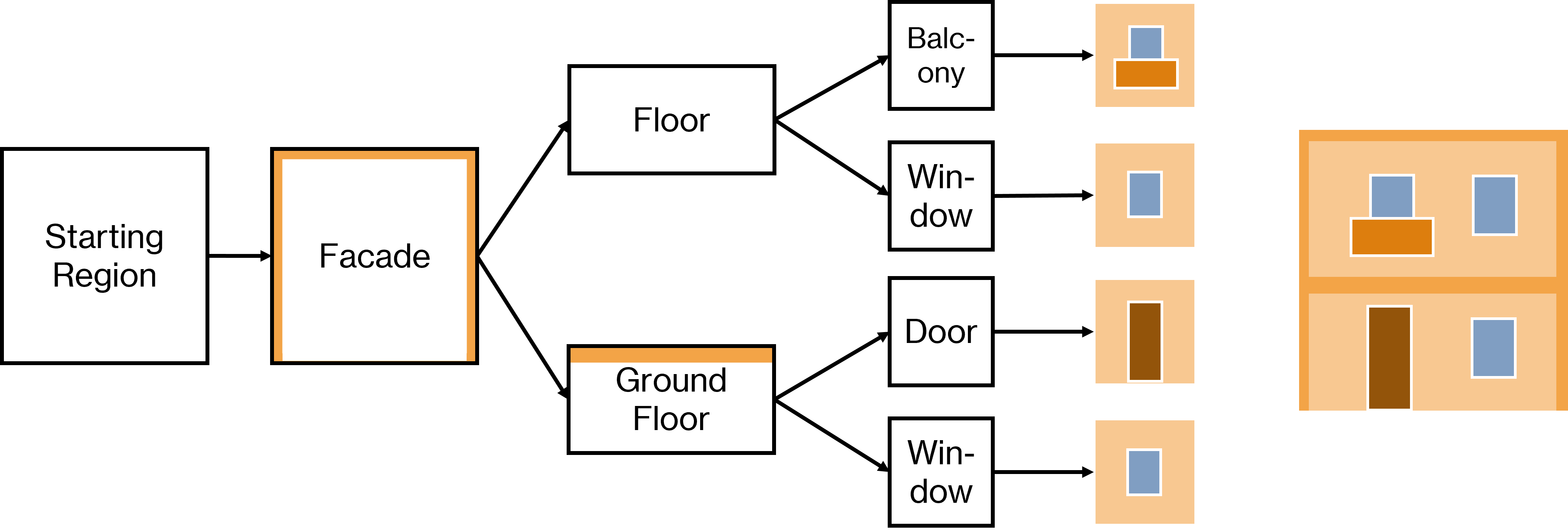}
        \caption{An example of a simplified procedural representation using a split grammar derivation tree. The tree defines the hierarchical structure of a facade and covers the whole image space.}
        \label{fig:procedure_example}
\end{figure}

A key aspect of our method is the hierarchical representation of facade structures. The structure of each facade, whether input or output, is defined by a split grammar derivation tree \citep{wonka2003instant}, forming a hierarchical procedural definition $P$ of the image space. Figure \ref{fig:procedure_example} illustrates a simplified example of such a representation. Using this representation, an edit can be defined as a pair of procedures:
$$
E = (P_{in}, P_{out}),
$$
where $P_{in}$ and $P_{out}$ are rooted trees of grammar split production rules which further and further divide the image space up until the leaf terminal symbols like walls, windows, etc.. Additionally, any intermediate nodes in the tree represent grammar nonterminals and still carry rich semantic information like "ground floor", "roof segment", etc. which is crucial for creating a mapping between the structures. By representing facades procedurally, we enable precise and interpretable edits to the structure, which is not achievable with pixel-based representations. This approach also enables hierarchy-driven diffusion model guidance, a novel contribution not present in the literature.

This approach is not limited to architecture. It can be adapted to other domains where hierarchical definitions over the image space are available.

\subsection{Procedural Reconstruction \& Editing}

\paragraph{Inverse Procedural Reconstruction.}
The method begins by extracting the procedural structure from the facade image $F_{in}$. To achieve this, we use Fa\c{c}AID \citep{facaid2024}, a transformer-based neuro-symbolic method for extracting procedural facade definitions from facade segmentations. This method  requires a facade segmentation as input to generate the procedural definition. This segmentation can be obtained automatically using state-of-the-art segmentation methods or created interactively by the user for more precise control \cite{coherance2012}.

The Fa\c{c}AID model outputs a procedural representation $P_{in}$ of $F_{in}$, which hierarchically divides the image space. This representation serves as the foundation for subsequent editing steps.

\paragraph{Structure Editing.}
After obtaining the structure, the user can modify the procedural definition to achieve the desired result. Edits can be performed in two ways:
\begin{enumerate}
    \item Adjusting parameters of the procedure, such as the number of floors or the sizes of windows.
    \item Modifying the hierarchy of the procedure to accommodate more complex edits, such as deleting every third balcony or adding more doors to the ground floor.
\end{enumerate}

\noindent
Regardless of the user's intended edit, the result should be a new procedure $P_{out}$ that remains valid within the constraints of the procedural language defined by the split grammar.

The pair of hierarchical  structure representations $(P_{in}, P_{out})$ defines the desired edit $E$ and serves as the foundation for the guidance scheme for the facade diffusion process. Pairing two semantically rich and hierarchical like that to form an edit enables precise and interpretable mapping between the two structures.

\subsection{Hierarchical Matching}
\label{subsec:matching}
The next step of the method involves creating a mapping between the original image space and the target image space. This mapping serves as the cornerstone of the guiding mechanism during the inference of the new image. Thanks to representing the input and target structures as grammar-based hierarchies, we can use the semantic information encoded in the trees to create this exact mapping.

The primary task now is to pair symbols from $P_{out}$ with corresponding source symbols from $P_{in}$. To achieve this, we developed a custom comparison metric comprising of two distinct components.

\paragraph{SVD Metric. }
A key goal of our metric is to compare the underlying structure of facade regions rather than simply evaluating pixel-level differences. For example, if a floor in the source structure has 4 windows and the target structure has 10 windows, even if the windows are identical and evenly spaced, pixel-based metrics like Mean Squared Error (MSE) would indicate a large discrepancy between the regions. To address this, we propose a new metric tailored for axis-aligned structured images, which aims to treat such regions as identical.

The foundation of our metric is Singular Value Decomposition (SVD). It is well-established in the literature that matrices with significant self-similarities and symmetries can be compactly represented as a sum of rank-1 matrices, also specifically for facade approximation \cite{yang2012rank1}. After performing SVD on a matrix $A$ as $\text{SVD}(A) = U^A, S^A, {V^A}^*$, we can construct an approximation $A_n$ by summing the first $n$ rank-1 matrices:
\[
A_n = \sum_{i=1}^{n} \sigma^A_i \cdot \mathbf{u}^A_i \otimes {\mathbf{v}^A_i}^T,
\]
where $\sigma^A_i$ is the $i$-th singular value, $\mathbf{u}^A_i$ is the $i$-th column of $U^A$, $\mathbf{v}^A_i$ is the $i$-th row of ${V^A}^*$, and $\otimes$ denotes the outer product. The more structured the matrix $A$, the smaller the value of $n$ required to accurately approximate it. The omitted singular values directly correlate with the MSE between $A$ and $A_n$:
\[
\text{MSE}(A, A_n) = \frac{1}{MN} \sum_{i=n+1}^{\min(M, N)} (\sigma^A_i)^2,
\]
where $M$ and $N$ are the dimensions of $A$.

\noindent
Using this relationship, we define the structural complexity of $A$ as
\[
\mathcal{C}_\epsilon(A) = \min \left\{ n \in \mathbb{N} \, \bigg| \, \frac{1}{MN} \sum_{i=n+1}^p (\sigma^A_i)^2 < \epsilon \right\},
\]
where $\epsilon$ is a predefined threshold that determines the acceptable level of the approximation. Two segmented image regions are considered structurally identical if their $\mathcal{C}_\epsilon$ values are the same.

To make this measure less discrete we additionally add a decimal part composed of the value of MSE normalized by $\epsilon$:
$$
\mathcal{C}'_\epsilon(A) = \mathcal{C}_\epsilon(A) + \frac{MSE(A, A_n)}{\epsilon}.
$$

\noindent
This allows us to fully define the structural difference metric of the matrices $A$ and $B$ as
$$
\mathcal{D}_{\text{SVD}}(A, B, \epsilon) = |\mathcal{C}'_\epsilon(A) - \mathcal{C}'_\epsilon(B)|.
$$

\paragraph{Histogram Metric. }
Additionally, we want to make the final metric aware of the contents of the regions being compared; otherwise, regions with the same structure but different terminals would still be treated as the same. That is why we also introduce an additional, content aware, custom metric $\mathcal{D}_{\text{H}}(A, B)$.
To compute the metric, we calculate histograms of intensity values in $A$ and $B$. We treat them as probability distributions, $p_A(i)$ and $p_B(i)$, and calculate the Hellinger distance between them:
$$
\mathcal{D}_{\text{H}}(A, B) = \sqrt{1 - \sum_{i} \sqrt{p_A(i) \cdot p_B(i)}}.
$$
In order to combine the SVD metric $M_{\text{SVD}}(A, B)$ and the histogram metric $M_H(A, B)$, we add their weighted values:
$$
\mathcal{D}(A, B, \epsilon) = \alpha \mathcal{D}_{\text{SVD}}(A, B, \epsilon) + \beta \mathcal{D}_{\text{H}}(A, B).
$$

\noindent
The pseudocode of the metric and its components is available in Appendix~\ref{appx:metric}.

\paragraph{Symbol Trees Mapping.}
Having constructed a suitable metric for comparing two regions in our hierarchical structures, pairing of grammar symbols in trees $P_{in}$ and $P_{out}$ can now be performed. Each node in the tree is augmented with a grayscale $r^i$ section of the facade segmentation occupied by each symbol $s^i$.

Starting from the symbol categories which appear closer to the root in the tree structures, for each symbol $s^i_{out}$ containing the region $r^i_{out}$ in $P_{out}$ we find the best matching symbol $s^i_{in}$ containing the region $r^i_{in}$ in $P_{in}$ by finding the pair producing the lowest value of the metric $\mathcal{D}(r^i_{in}, r^i_{out}, \epsilon)$. An important additional restriction is that when a parent of a symbol has already been matched, the available choices for the hierarchical matching are reduced to only the children of the matched parent.

Given a function $r(s)$, which for a given symbol returns its corresponding region, and a function $\operatorname{cat}(s)$, which for a given symbol returns its category, we can define the hierarchical matching mathematically as a function $M(s)$ mapping output symbols to input symbols in the following manner: 
\[
M\!\left(s^{i}_{\mathrm{out}}\right)
=
\operatorname*{argmin}_{\,s\in \Omega(s^{i}_{\mathrm{out}})}
\mathcal{D}\!\left(r(s),\, r(s^{i}_{\mathrm{out}}),\, \epsilon\right), \;\;\text{where}
\]
\[
\Omega(s^{i}_{\mathrm{out}})=
\begin{cases}
\{\,s\in P_{\mathrm{in}}\;:\;\operatorname{cat}(s)=\operatorname{cat}(s^{i}_{\mathrm{out}})\,\}, 
\\\text{if } \operatorname{parent}(s^{i}_{\mathrm{out}})\ \text{is not matched},\\[6pt]
\{\,s\in \operatorname{children}\!\big(M(s^{i}_{\mathrm{out}})\big)\;:\;\operatorname{cat}(s)=\operatorname{cat}(s^{i}_{\mathrm{out}})\,\},
\\\text{otherwise.}
\end{cases}
\]
In practice, computing this mapping produces a list of paired symbols, and subsequently regions in the two image spaces. This provides an educated semantically-driven correlation between the two facade structures and can be used later in the pipeline to provide meaningful guidance.

The pseudocode for the full hierarchical matching algorithm is available in Appendix~\ref{appx:matching}.

\begin{figure}[t]
\captionsetup[subfigure]{labelformat=empty}
    \centering
    \input{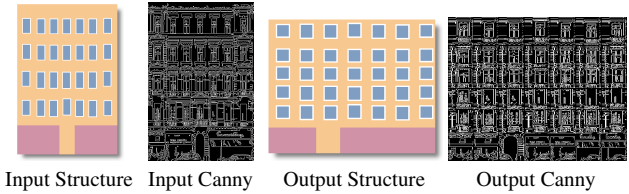}
    \caption{Example of the fully reconstructed Canny edges serve as guidance during the inference process. The new Canny edges image is created by transforming the original Canny edges image according to the hierarchical pairings.}
    \label{fig:canny_construction}
\end{figure}

\subsection{Guided Inference}

Having constructed the mapping between the two image spaces, the next step is to use this information in a guided diffusion setting.

\paragraph{Null-text Inversion.}
We first perform Null-Text Inversion \citep{nulltext2023} on the input facade $F_{in}$, reconstructing it as a diffusion model output and capturing the input noise needed to approximate $F_{in}$. We also save the network activations $\Psi_{in}$ used to generate $F_{in}$, which provide semantic information for subsequent edits. Since the ControlNet we use is conditioned on Canny edges, we simultaneously compute Canny edges $C_{in}$ from $F_{in}$. The whole inversion step is done once per input facade and can be reused for multiple edits.

\paragraph{Guiding Input Construction.}
Next, we construct two guidance inputs: (1) a new Canny edges image $C_{out}$ and (2) new target activations $\Psi_{out}$. For each terminal region pair from the hierarchical matching, we copy and resize segments from $C_{in}$ and $\Psi_{in}$ to form $C_{out}$ (Figure~\ref{fig:canny_construction}) and $\Psi_{out}$. Linear interpolation is applied whenever source and target sizes differ.

\paragraph{Optimization.}
Finally, we run a guided inference pass, using the extracted noise as the starting point and $C_{out}$ as conditioning for ControlNet to match the desired structure. We also follow \citet{pandey2024diffusionhandles} to optimize the latent image toward $\Psi_{out}$. Specifically, we minimize an energy function that penalizes the $L_2$ distance between current activations $\Psi'$ and $\Psi_{out}$ under a curated weight schedule. After the full diffusion process, decoding yields the edited facade $F_{out}$, reflecting the procedural edit $E=(P_{in}, P_{out})$ applied to $F_{in}$.

\begin{figure}[t]
\captionsetup[subfigure]{labelformat=empty}
    \centering
    \input{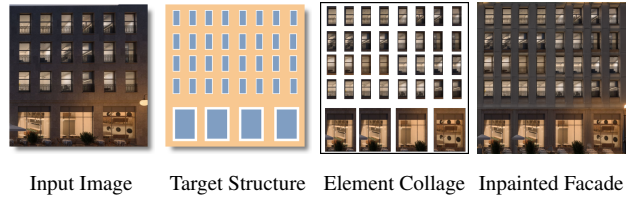}
    \caption{Visualization of the Photoshop inpainting baseline creation process. A collage of facade elements is assembled, followed by inpainting to complete the remaining portions of the facade.}
    \label{fig:bl_photoshop}
\end{figure}

\section{Evaluation}\label{sec:evaluation}
In this section, we perform an extensive evaluation of our pipeline on a set of $50$ real-word facade images. \begin{revision} We utilize a representative subset of 20 facades to provide a qualitative analysis and to anchor our user study. To ensure the statistical rigor of our findings, we perform the quantitative evaluation across the complete set of 50 facades. \end{revision}

\subsection{Qualitative Results}
The results of our pipeline are depicted in Figure~\ref{fig:qualitative_results}, showcasing generation results for 20 different facades, each with two distinct user-driven edits. Each entry consists of five images: (1) the original facade image, (2) the target structure for edit no. 1, (3) the inference result for edit no. 1, (4) the target structure for edit no. 2, and (5) the inference result for edit no. 2. The target structure visualizations showcased in the Figure are composed of aggregated leaves (terminals) from the grammar-based hierarchical tree inputs.

The results demonstrate that our approach successfully generates edited versions of facades across various structures and styles while preserving the core identity of the original design, making the new facade perceptually look like an edited version of the input. The output images closely match the target structures and achieve a level of realism comparable to typical outputs of the underlying model used in the pipeline (Stable Diffusion 1.5).

\subsection{Baselines}
\label{subsec:baselines}
Traditional editing methods for diffusion-based image editing typically target operations on individual elements \cite{pandey2024diffusionhandles, song2023objectstitch} while keeping the rest of the image mostly unchanged. Our method, however,  aims to fully reconstruct the whole image space in one go, while at the same time usually also changing the counts of the individual elements between the input and target designs. This level of structural control over the edit makes the other state-of-the-art editing approaches unusable for performing the task. This means that to provide any means of a fair comparison, we needed to develop custom baselines, based on currently available tools, that more accurately match the parameters of our full-image-restructuring approach.

\begin{figure*}[h]
    \centering
    
\newlength{\imgw}    \setlength{\imgw}{0.135\textwidth}
\newlength{\gap}     \setlength{\gap}{1.45em}
\newlength{\arroww}  \setlength{\arroww}{1.9em}
\newlength{\rowgap}  \setlength{\rowgap}{0.6em}
\newlength{\headgap} \setlength{\headgap}{0.4em}

\newcolumntype{C}[1]{>{\centering\arraybackslash}m{#1}}
\newcolumntype{A}{>{\centering\arraybackslash}m{\arroww}}

\newcommand{\colrule}{\hspace{\gap}\vrule width 0.6pt\hspace{\gap}}

\newcommand{\arrowcell}{\scalebox{1.8}{$\boldsymbol{\Rightarrow}$}}

\newcommand{\img}[2][0 0 0 0]{%
  \includegraphics[width=\linewidth,keepaspectratio,trim=#1,clip]{#2}%
}

\newcommand{\imagerow}[5]{%
  \img{#1} & \arrowcell & \img{#2} & \img{#3} & \img{#4} & \img{#5} \\
  \addlinespace[\rowgap]
}
\noindent\centering
\begin{tabular}{
  @{} C{\imgw} @{\hspace{\gap}} %
      A        @{\hspace{\gap}} %
      C{\imgw} !{\colrule}      %
      C{\imgw}  C{\imgw}  C{\imgw} @{}
}
\textbf{Input} & & \textbf{Target} & \textbf{ControlNet} & \textbf{Photoshop} & \textbf{Ours} \\
\addlinespace[\headgap]

\imagerow{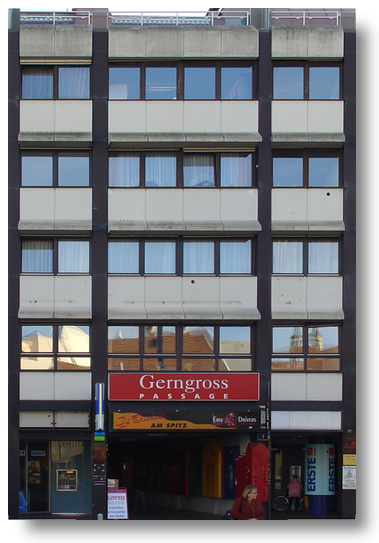}
          {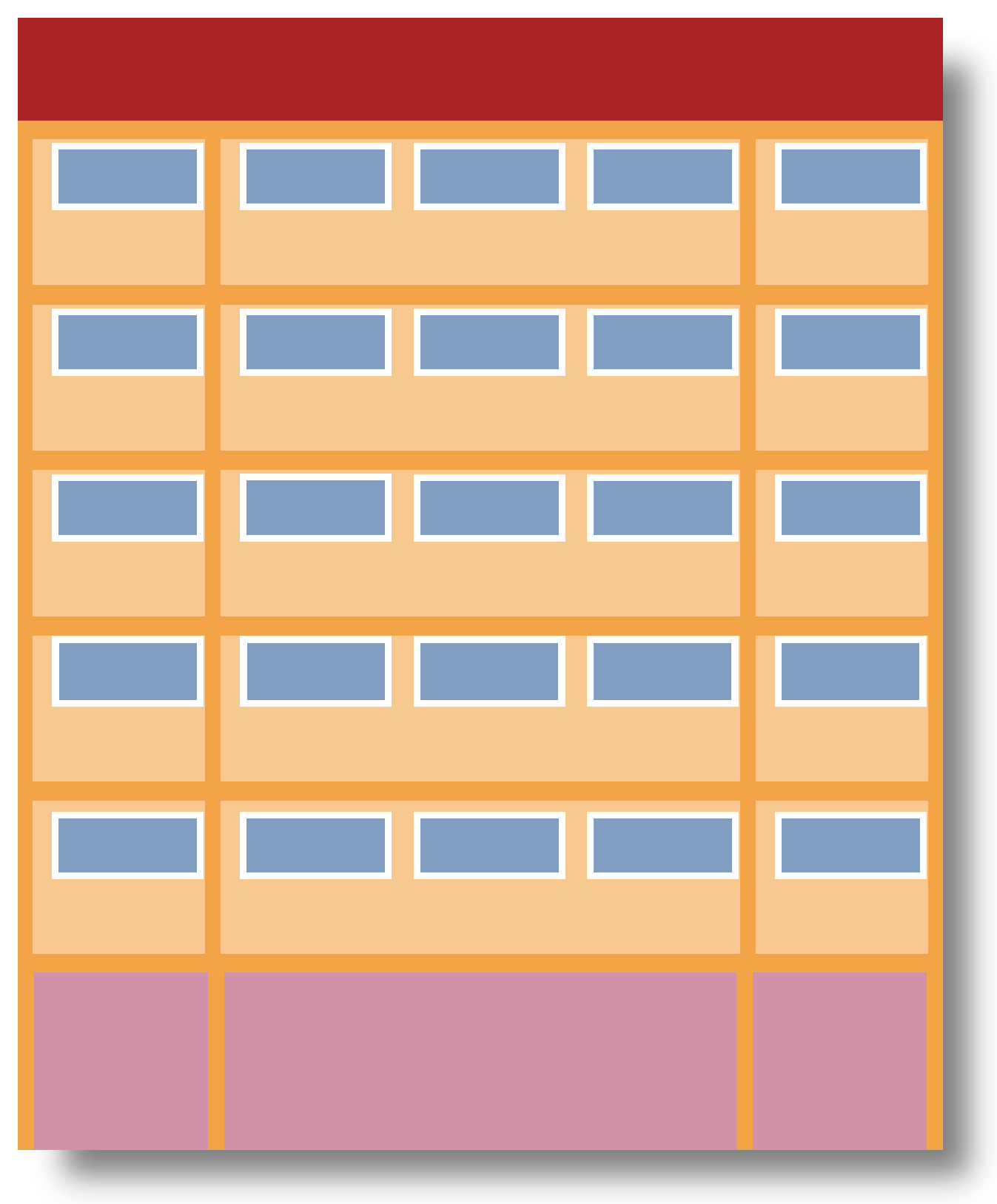}
          {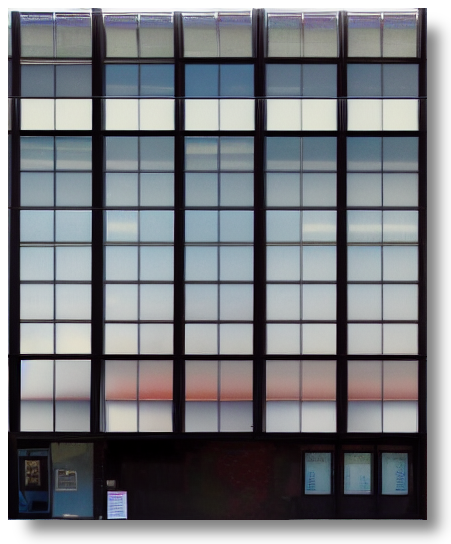}
          {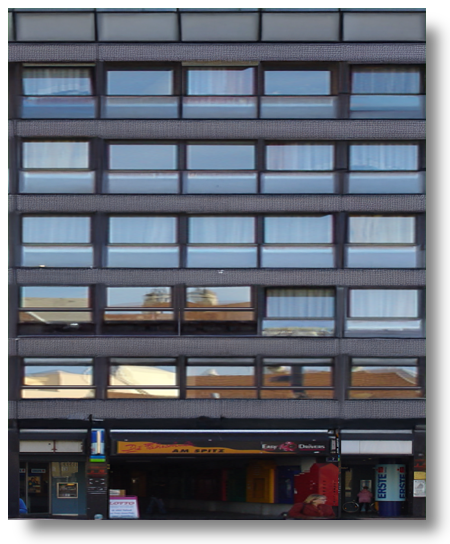}
          {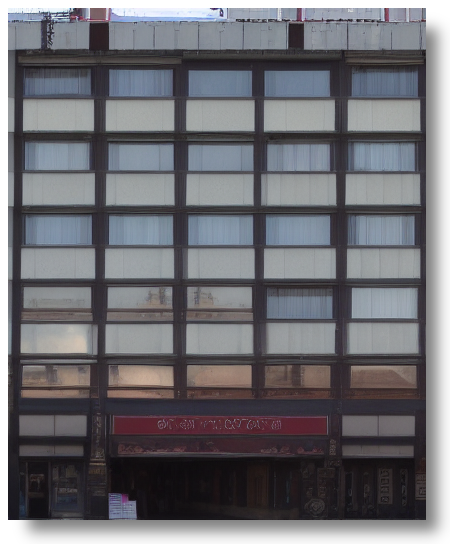}

\imagerow{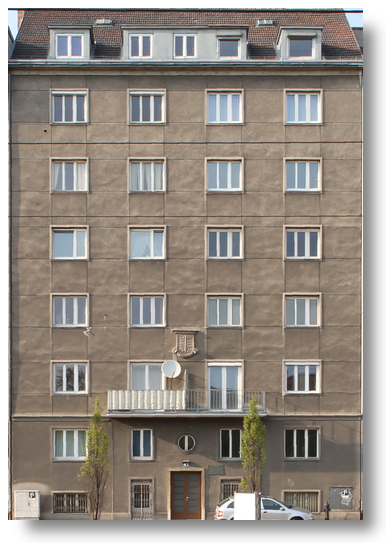}
          {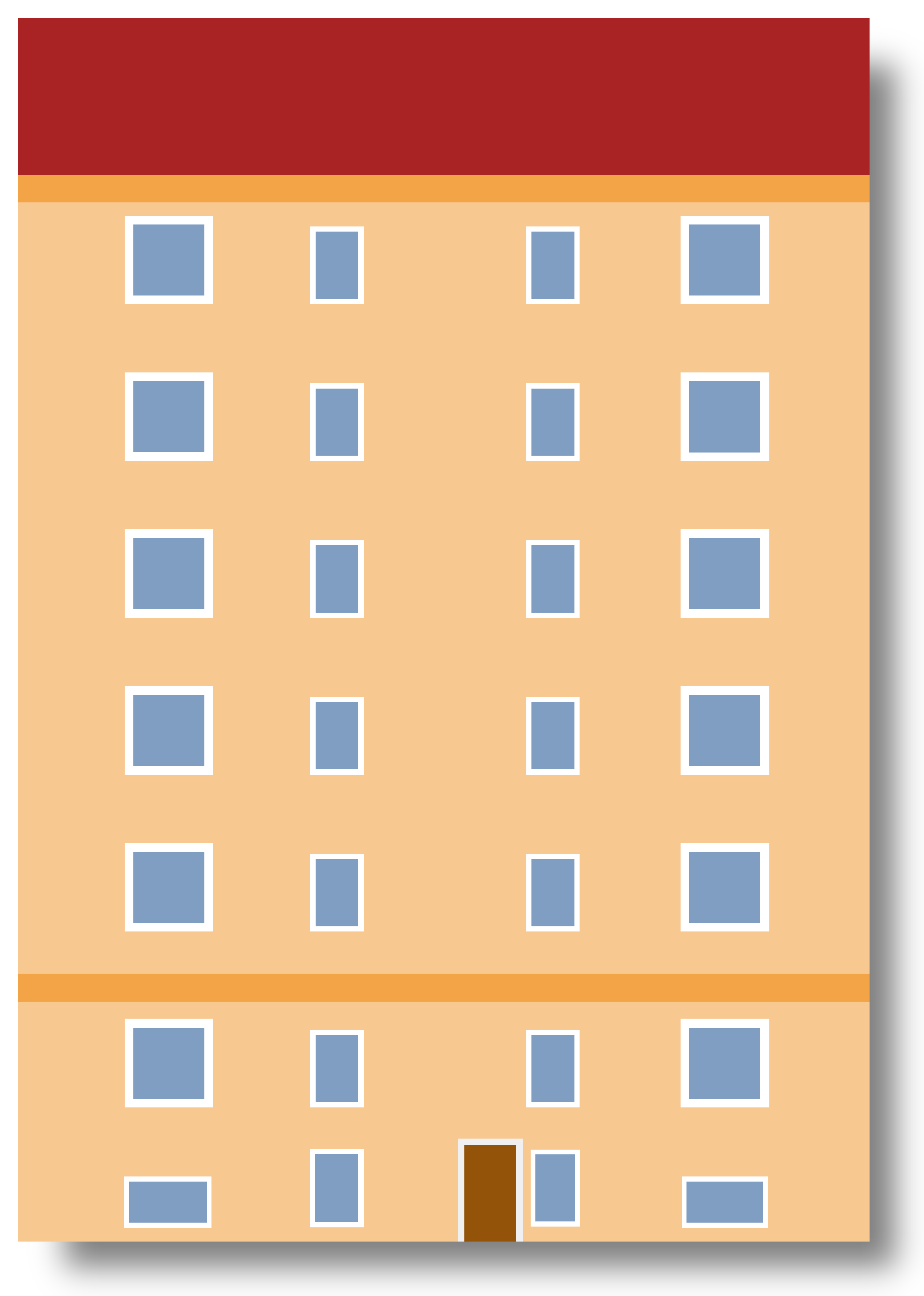}
          {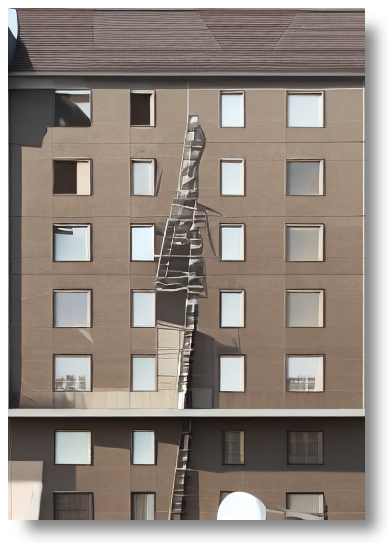}
          {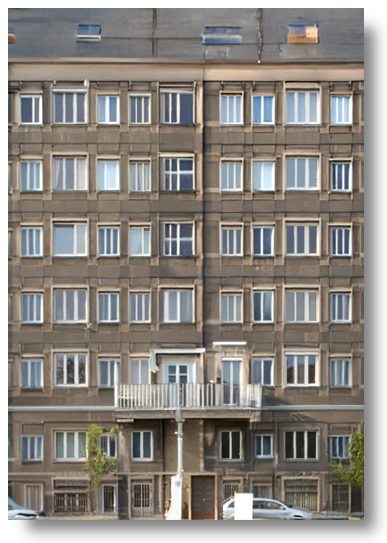}
          {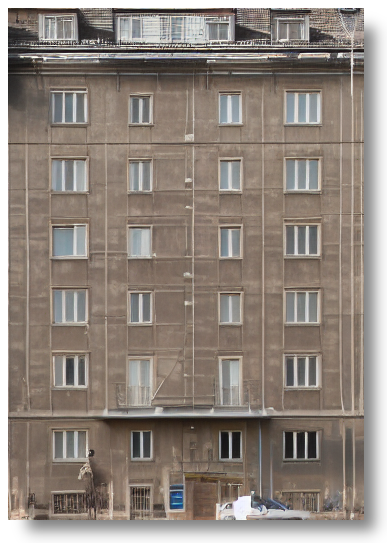}

\imagerow{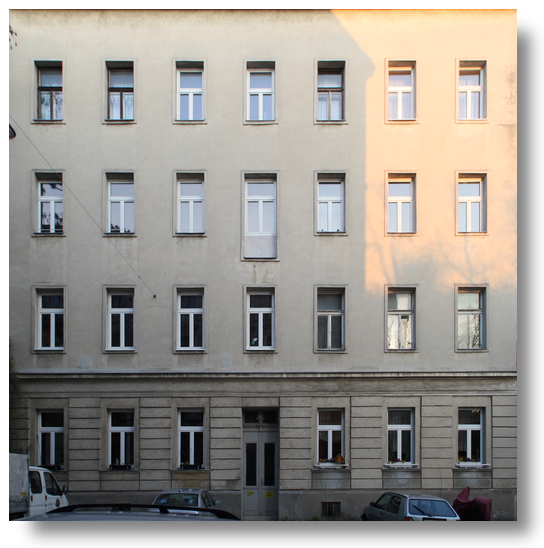}
          {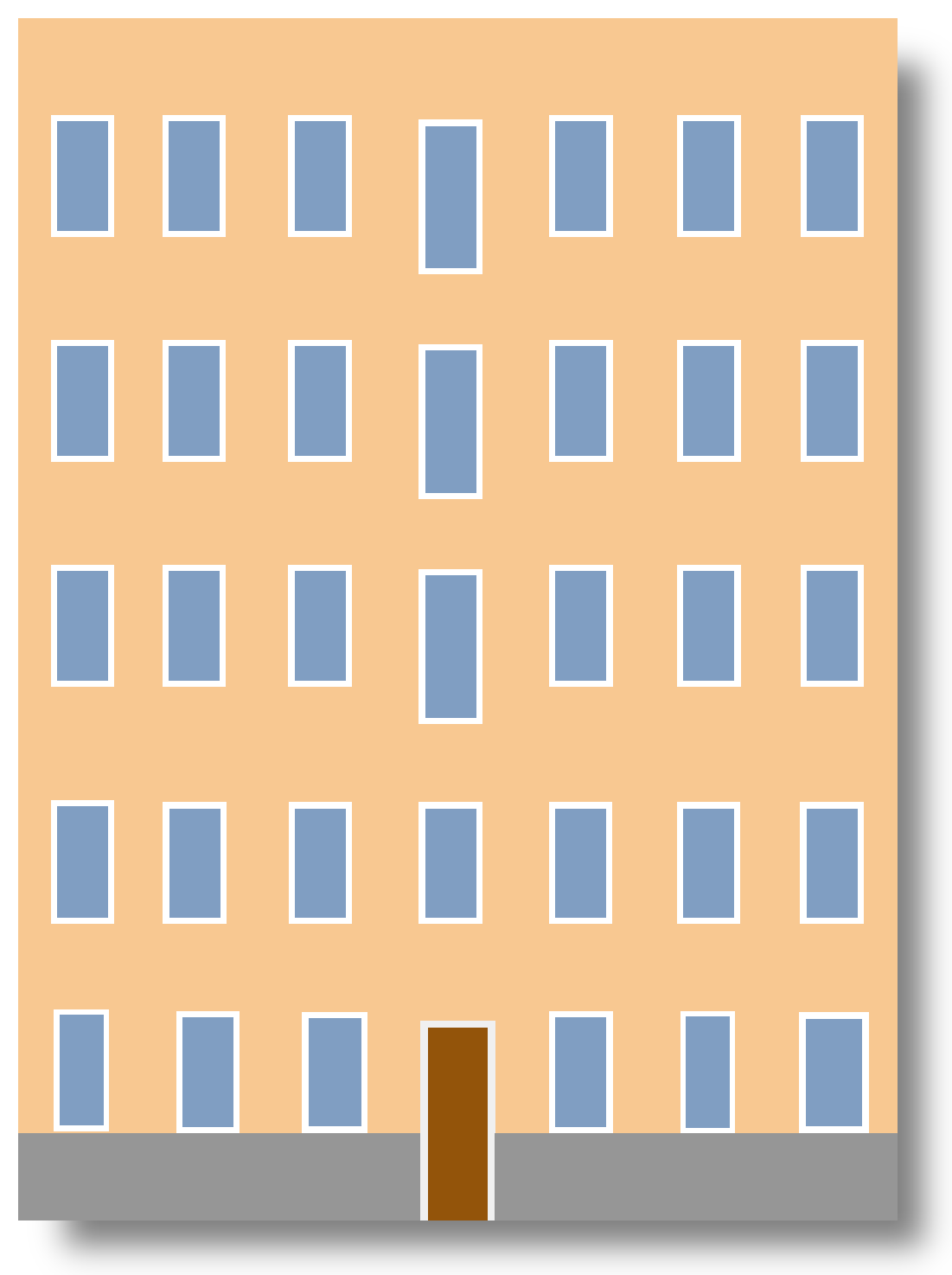}
          {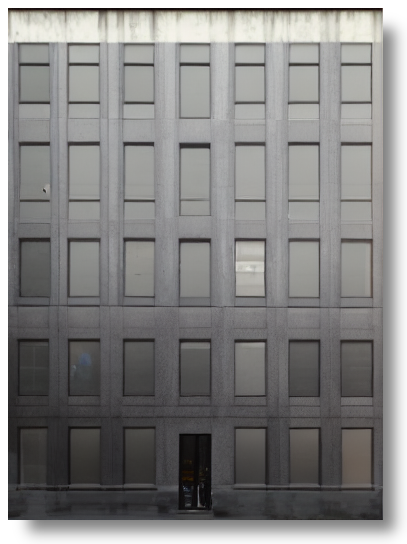}
          {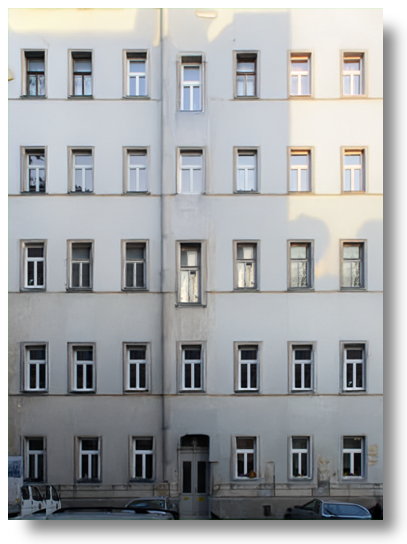}
          {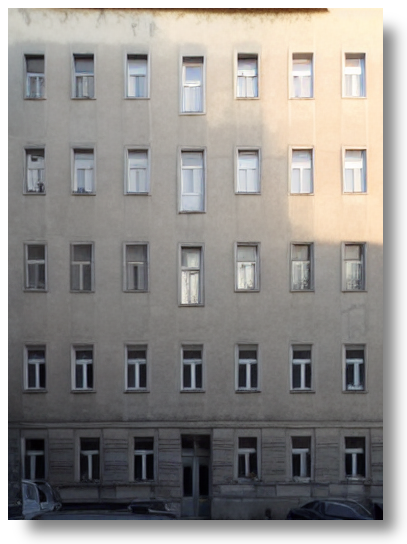}

\imagerow{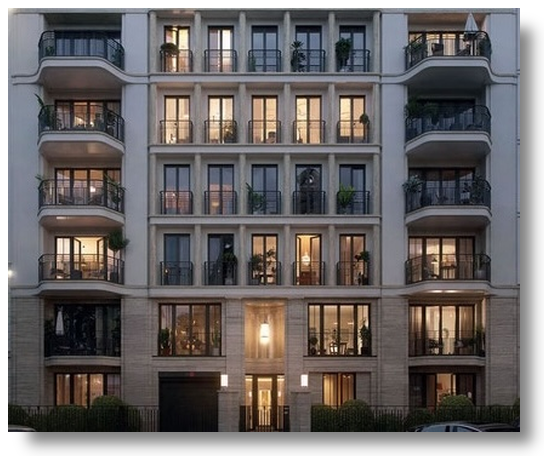}
          {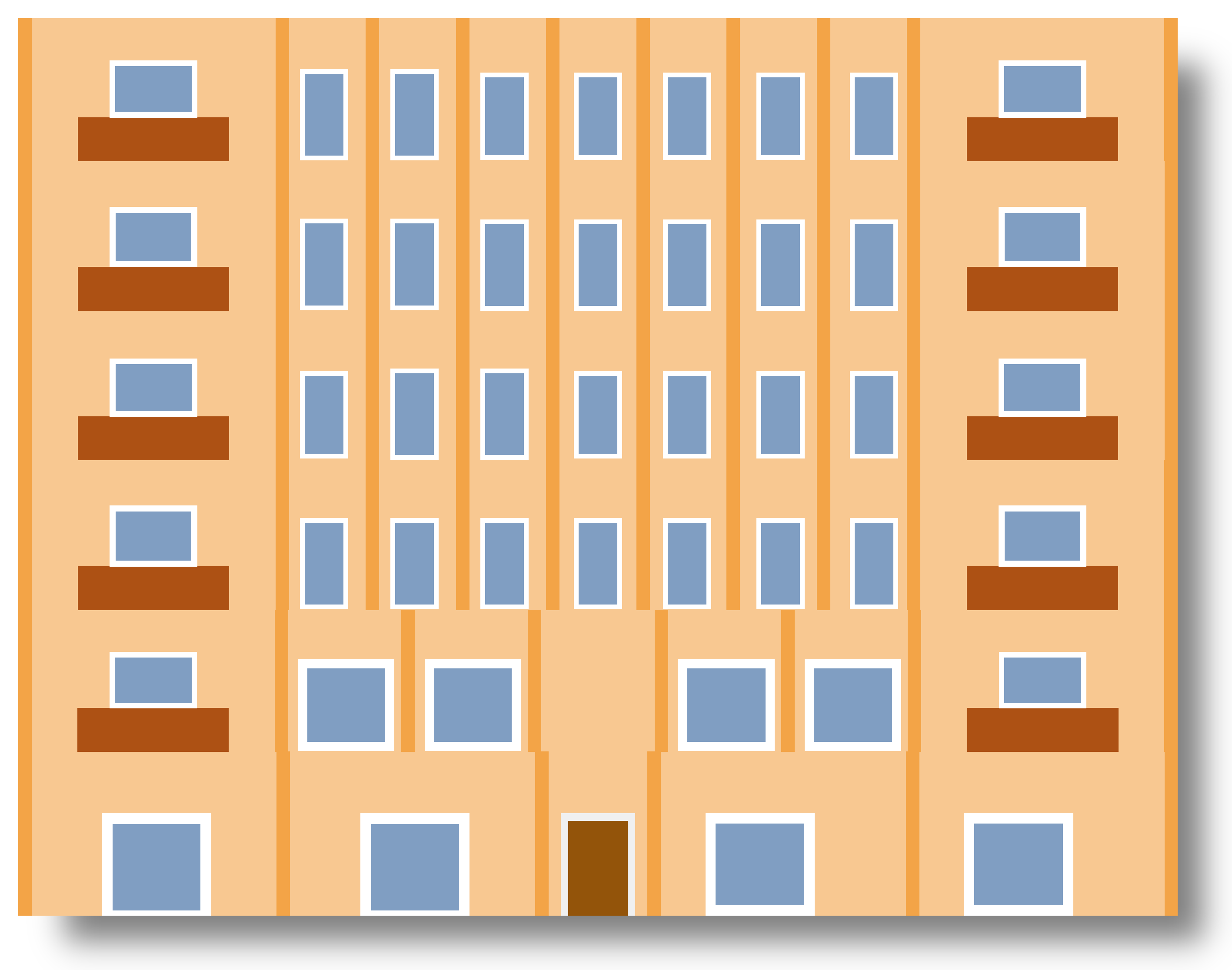}
          {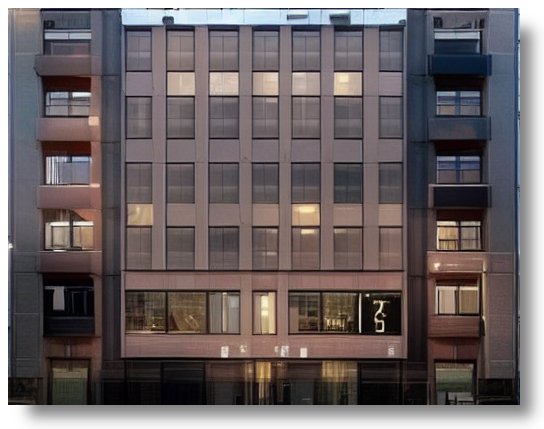}
          {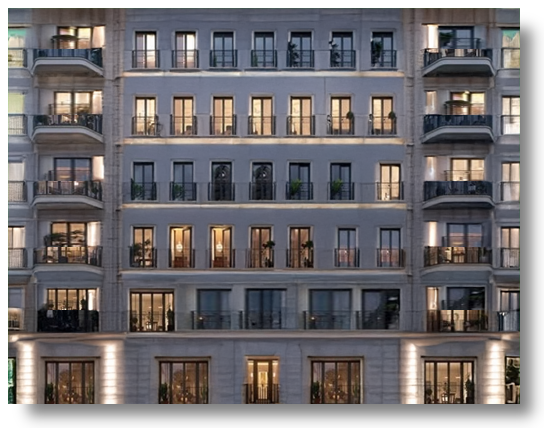}
          {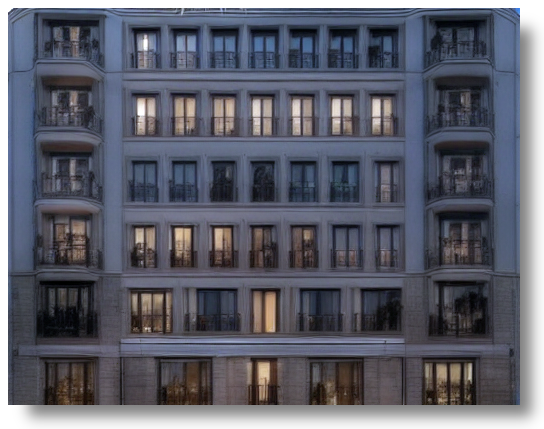}

\imagerow{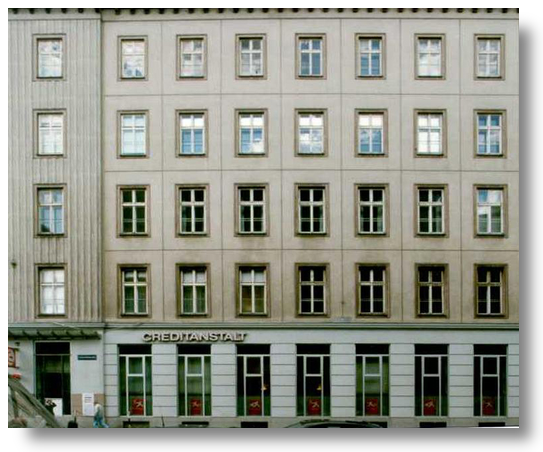}
          {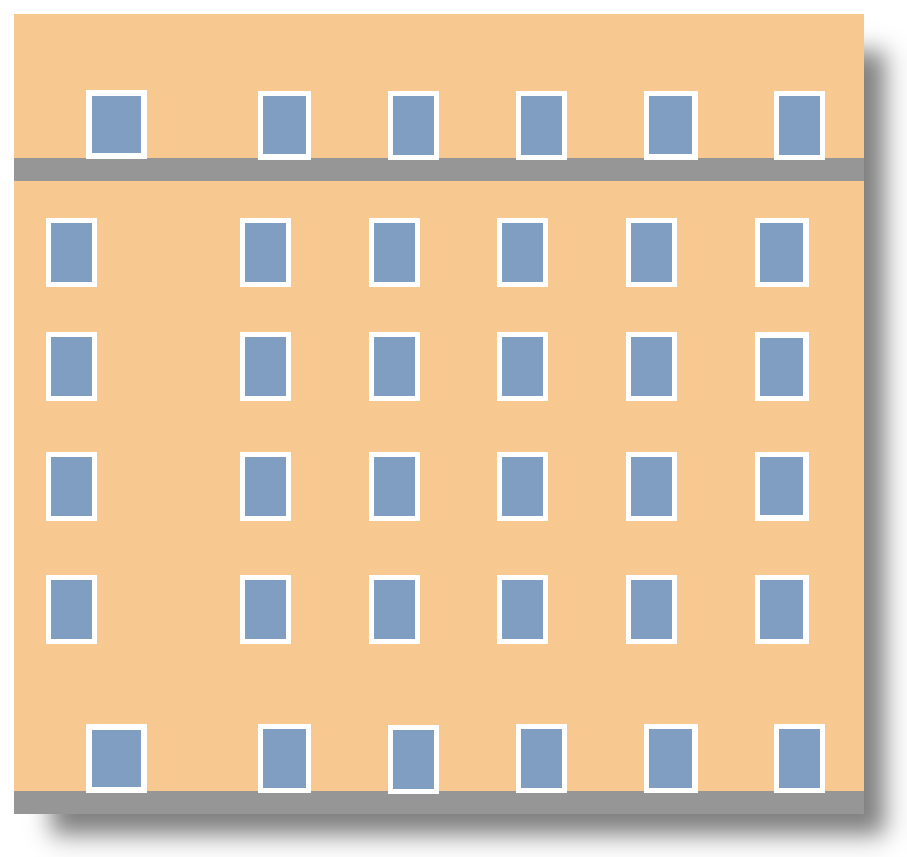}
          {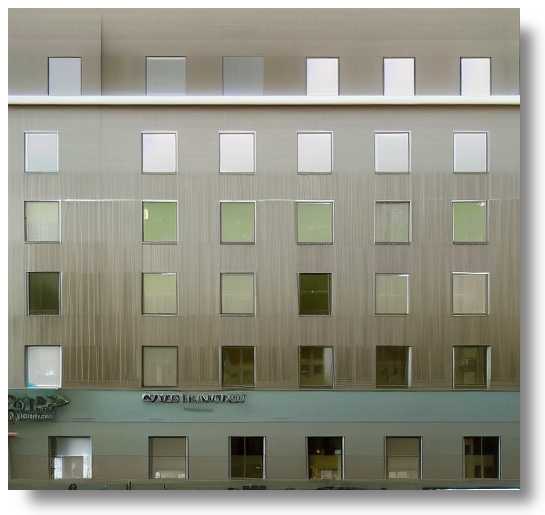}
          {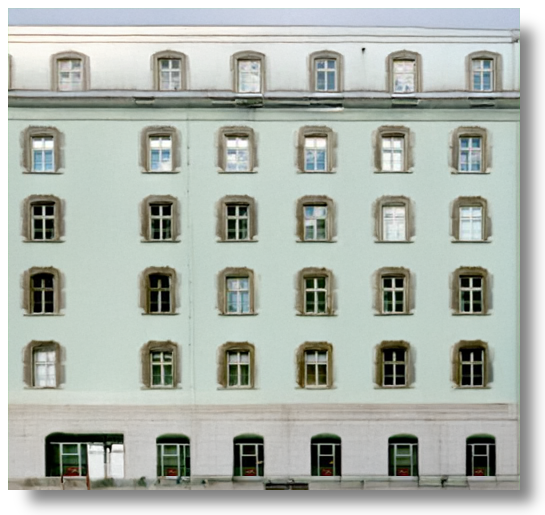}
          {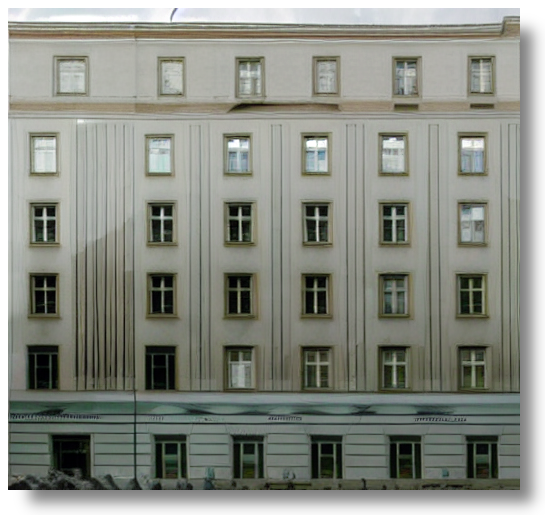}

\imagerow{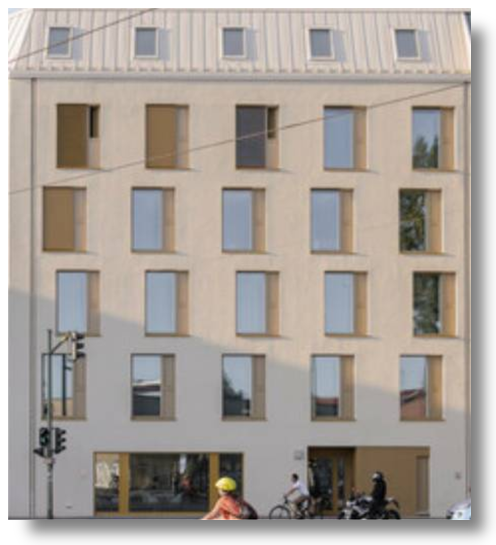}
          {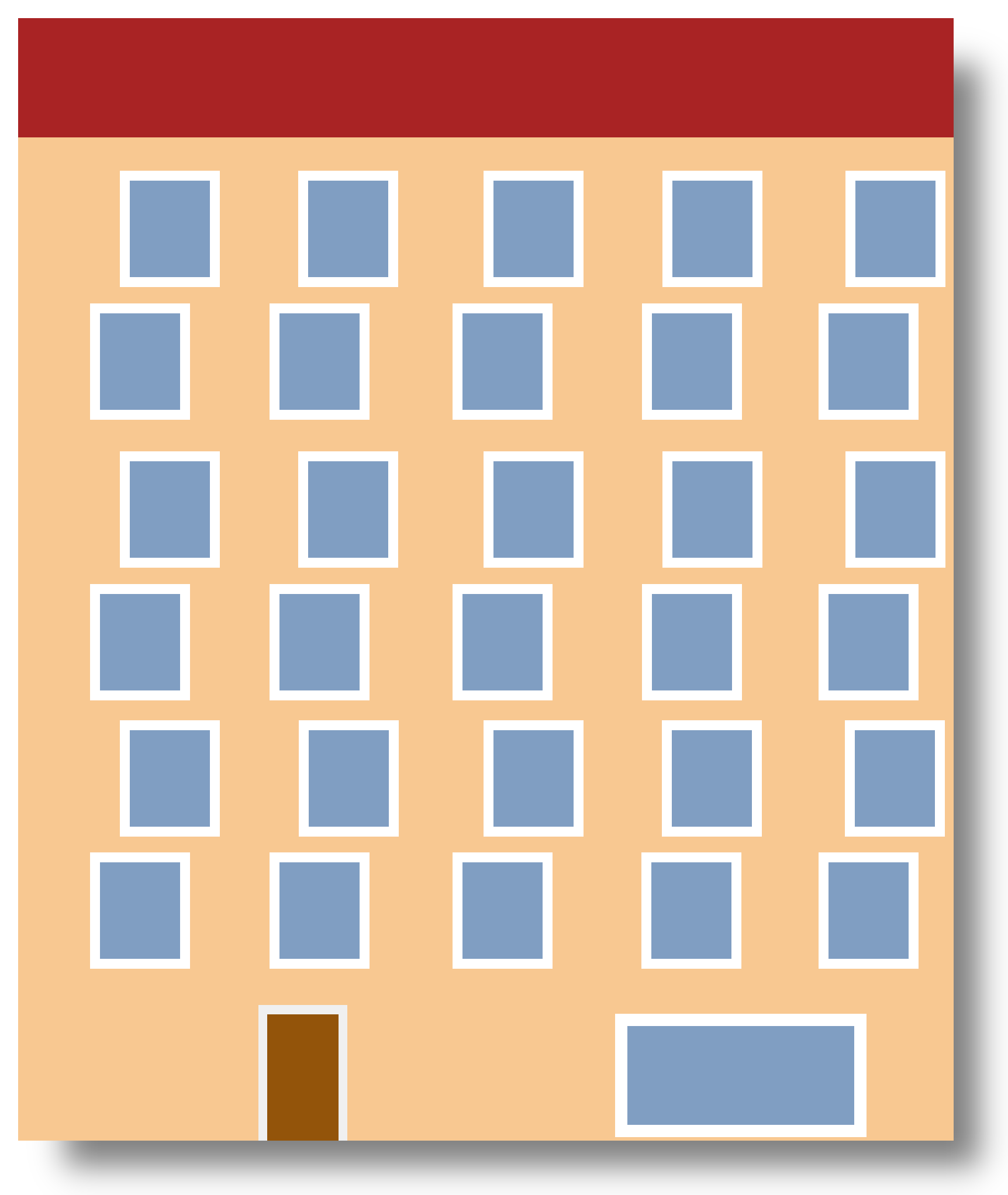}
          {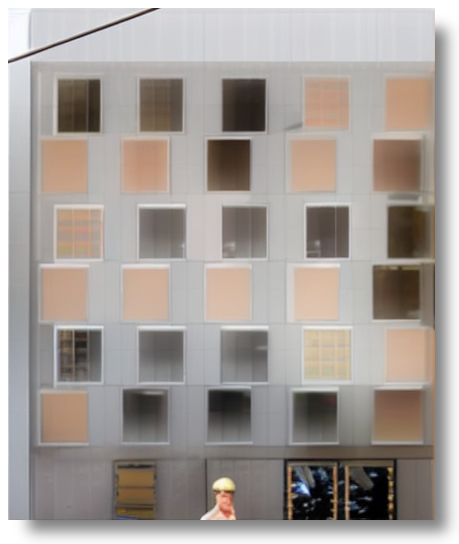}
          {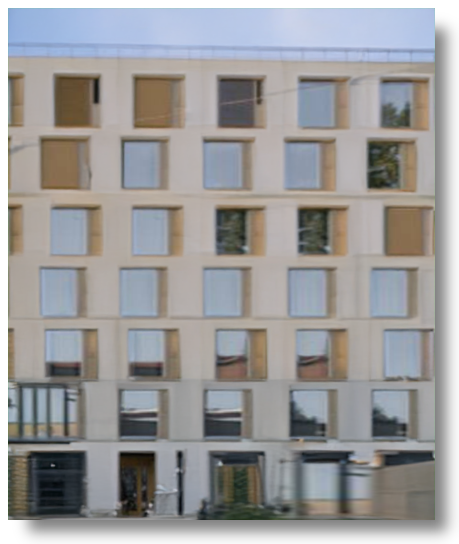}
          {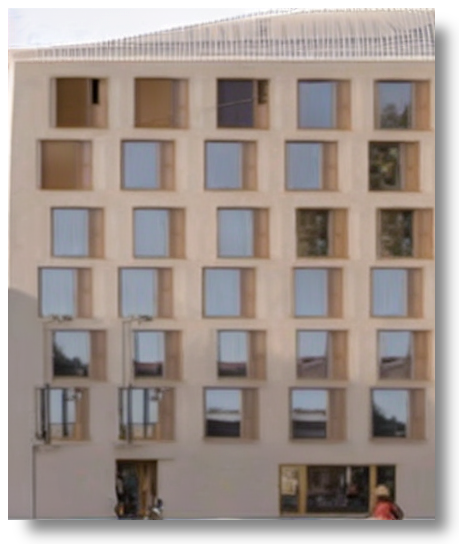}

\imagerow{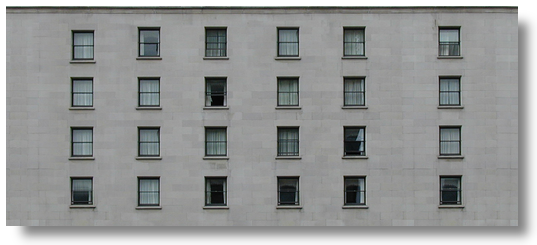}
          {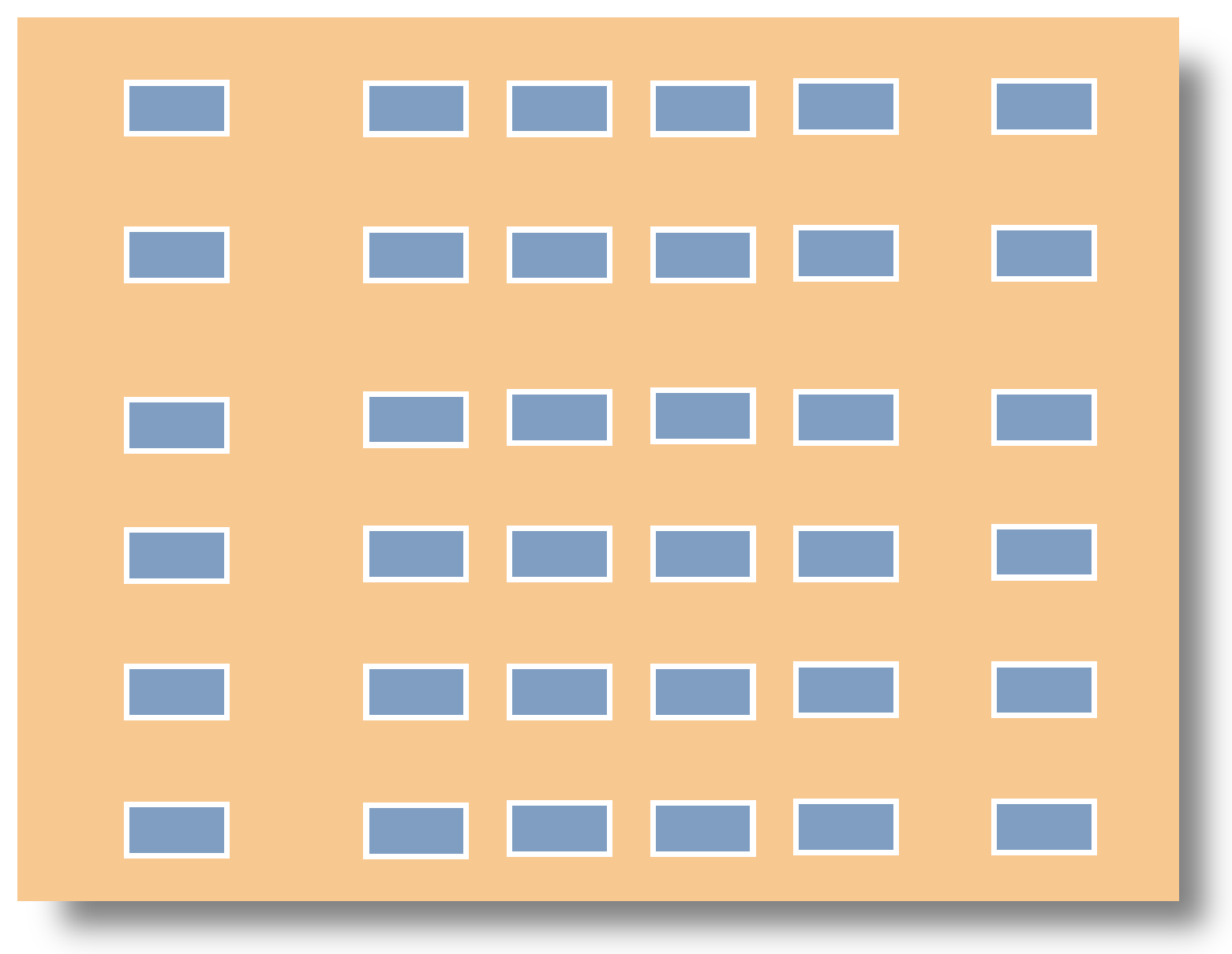}
          {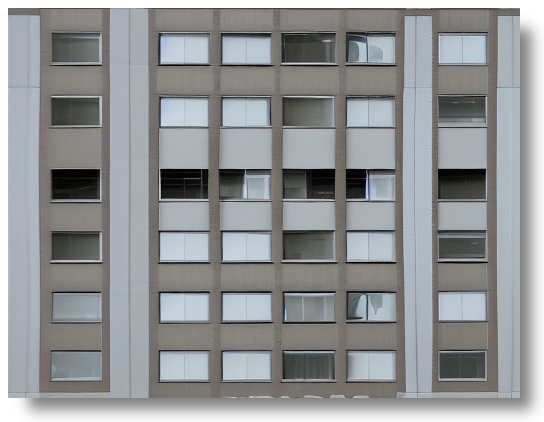}
          {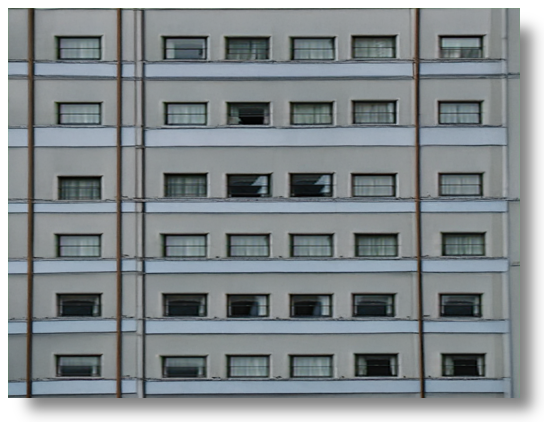}
          {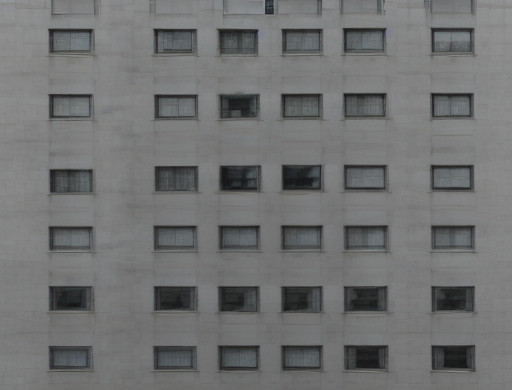}

\end{tabular}

  \caption{Comparison to baselines. Each row starts with the input image, followed by the target, a ControlNet baseline result, a Photoshop baseline result, and the result of our method.}
    \label{fig:comparison_results}
\end{figure*}

\paragraph{ControlNet Baseline.}
Since our hierarchical grammar-based definitions can easily be turned into segmentations, the natural approach to generate new images with a given segmentation with a diffusion model would be to use a standard segmentation based version of ControlNet. However, this would generate a brand new facade design unrelated to the input image, instead of editing it. Having that in mind we construct the ControlNet Baseline by first performing Null-text Inversion on the input image using the segmentation ControlNet model (with the input segmentation passed as the control structure) to extract the starting noise which should contain indirectly encoded information about the original design. We then perform inference starting from this noise but swamping the control structure to the target segmentation. The results of this approach are showcased in Figure~\ref{fig:comparison_results} in the ControlNet column.

\paragraph{Photoshop Baseline.}
We construct another baseline by adopting a collage-based inpainting approach. First, elements from the original facade image, like windows and doors, are repositioned and rescaled to match the target structure. From our testing, creating a collage like that manually takes about 10 minutes for a moderately experienced image editing software user. To allow for easily generating more results, we streamline the process by augmenting this baseline with our hierarchical matching algorithm. After generating the image space mapping based on the grammar-based representations, we can automatically generate a collage by cropping specific elements from the input image and pasting them onto the collage. While this baseline approach can function on its own by sticking to the manual 10 minute edits, in order to generate more examples, we needed to let it borrow the hierarchical matching part of our pipeline.

We then use Adobe Photoshop’s~\cite{adobe2025photoshop} generative fill tool, Adobe Firefly\cite{adobe2023firefly}, to in-paint the remaining image space (Figure~\ref{fig:bl_photoshop}). We generate the collages with three different padding settings to test which one provides the best wall context. From all padding levels and all resulting variants generated by Firefly, we select the best looking version for each edit. The results of this approach are showcased in Figure~\ref{fig:comparison_results} in the Photoshop column.

\paragraph{Renderings Baseline.}
While the inputs to our method are real facade images, the expected resulting designs are differently structured hypothetical version of said facade. This means that a typical ground truth does not exist.
To establish a reliable ground truth, we constructed 3D models of some of the input facades using specific assets—windows, doors, balconies—positioned according to the grammar-based definitions. This process produced synthetic versions of the original facade designs and allowed flexible manipulation in 3D. By repositioning and transforming these assets to match the edit target structure, we obtain hypothetical “ground truth” renderings illustrating the expected facade designs after performing the edits.

\begin{figure}[t]
\centering
\input{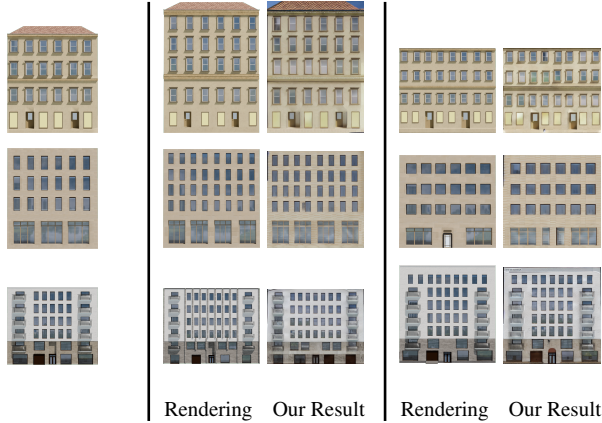}
\caption{Comparison between the renderings of modified facade models and the corresponding results from our method on the same facade structures. Each row corresponds to one facade, with Input on the left, Edit 1 in the middle, and Edit 2 on the right.}
\label{fig:bl_renderings}
\end{figure}

We run the initial renderings through our pipeline to see how well the output edited images match the expected hypothetical "ground truth" renderings. The results (Figure~\ref{fig:bl_renderings}) closely resemble the intended structures and designs; minor textural artifacts appear, likely because standard diffusion models are not trained on fully-lit, semi-realistic 3D renderings.

\subsection{User Study}
\label{par:user_study}
To evaluate the quality of the generated results, we conducted a user study involving all 40 edits (2 per facade). The study aimed to measure three key aspects of the resulting images:
\begin{itemize}
\item \textbf{Realism.} Does the generated image look like a plausible facade?
\item\textbf{Appearance Preservation.} Does the new facade retain the core appearance of the original design?
\item \textbf{Edit Adherence.} Does the new facade align with the target procedural representation?
\end{itemize}

\noindent
For comparison, we generated the same edits using the previously described ControlNet and Photoshop baselines. We have run a separate study for each of the baselines.

During the study, participants were presented with pairs of facades and asked to select the better-performing image for each aspect (they could also deem their performance equal). To provide additional context, the original facade image and the target structure were also displayed. A total of 105 unique users participated, evaluating 1015 facade pairs---answering 3045 questions in total. Appendix~\ref{appx:userstudy} provides more details regarding how the study was conducted.

\begin{figure}[t]
    \centering
    \begin{subfigure}[b]{0.48\columnwidth}
        \centering
        \includegraphics[width=\linewidth]{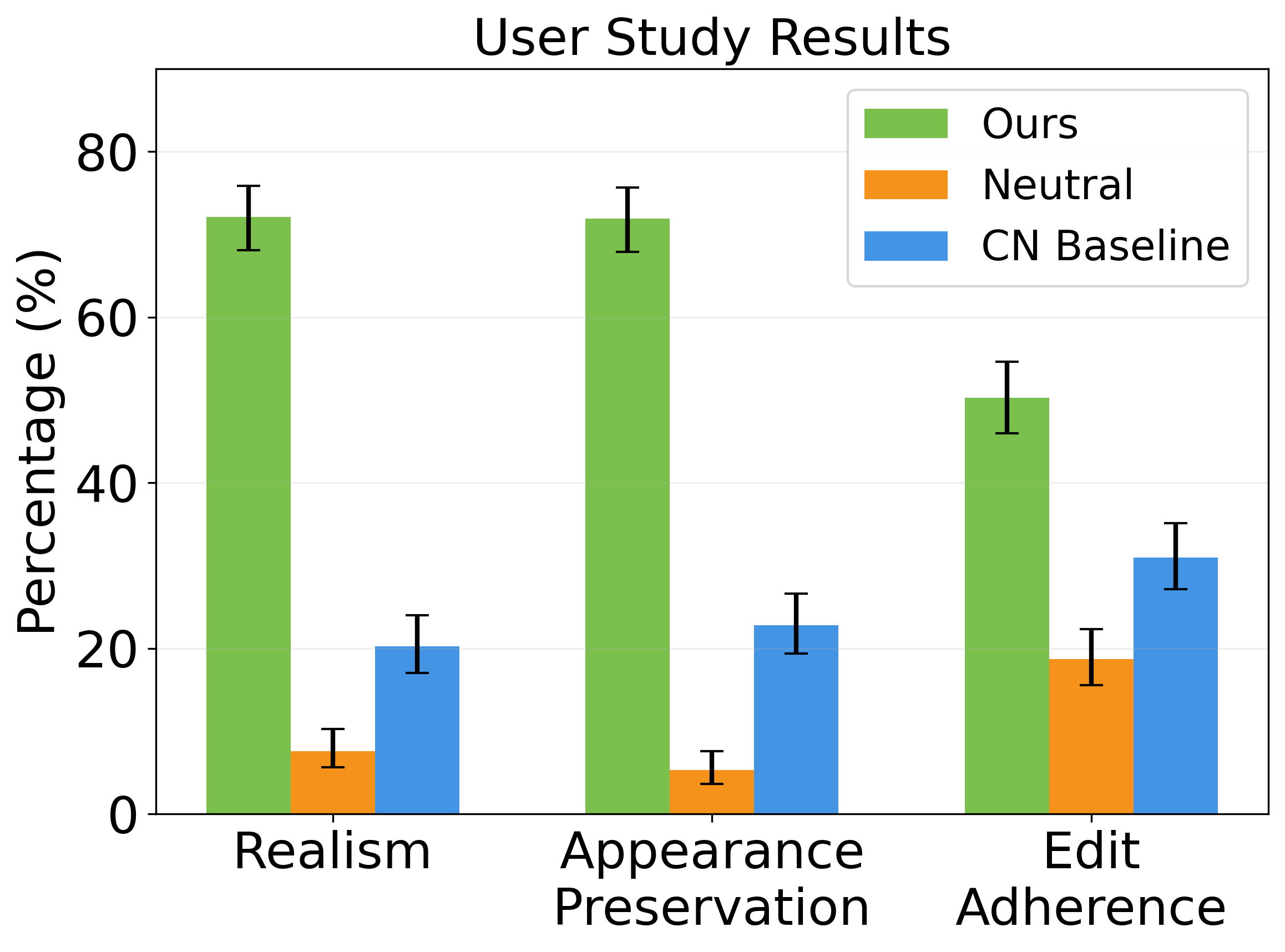}
        \caption{ControlNet Baseline}
        \label{fig:user_study_a}
    \end{subfigure}
    \hfill
    \begin{subfigure}[b]{0.48\columnwidth}
        \centering
        \includegraphics[width=\linewidth]{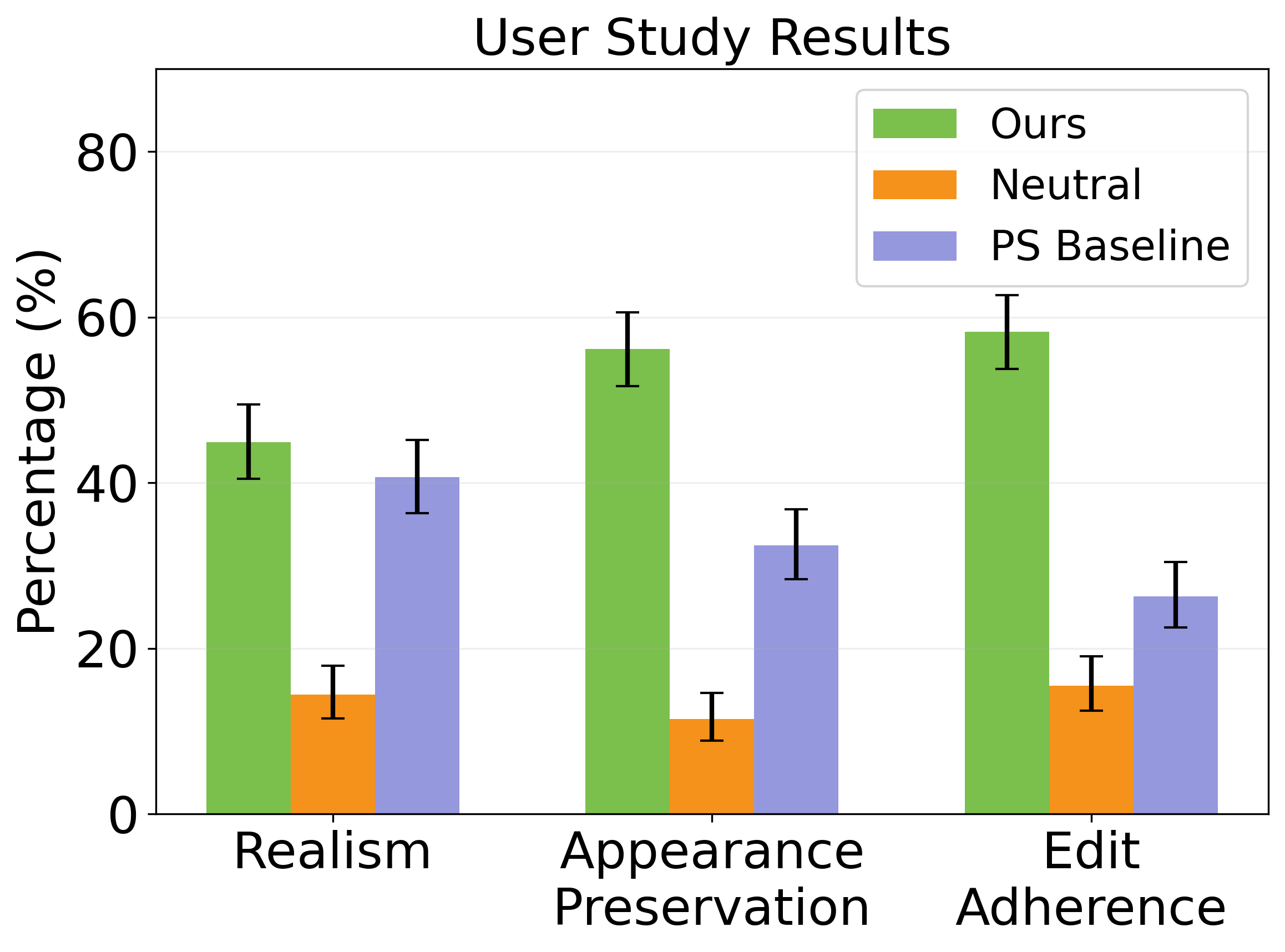}
        \caption{Photoshop Baseline}
        \label{fig:user_study_b}
    \end{subfigure}

    \caption{User study results showing if the users were partial to our results (Ours), to the baseline (CN/PS Baseline), or if they had no preference (Neutral). Results are separated into three categories corresponding to the questions asked.}
    \label{fig:user_study}
\end{figure}

\paragraph{ControlNet Baseline.} The vast majority of users preferred the results of our method when compared to the segmentation based ControlNet baseline when it comes to realism and appearance preservation (Figure~\ref{fig:user_study_a}). The only metric where the split was closer---but still favored our edits---was the edit adherence. This is to be expected since ControlNet was specifically designed to match a given control structure.

\paragraph{Photoshop Baseline.} The results (Figure~\ref{fig:user_study_b}) show that even though the Photoshop baselines were created through a semi-automatic, curated process and employed the help of our hierarchical matching algorithm, our pipeline outperformed the Photoshop approach in both the edit adherence and identity preservation. Additionally, even though our method uses a significantly less complex foundational model (Stable Diffusion 1.5), than the generative fill function in Photoshop (Adobe Firefly), there was no statistically significant difference between the two approaches when it comes to perceived realism.

\subsection{Quantitative Evaluation}\label{sec:quantitative}

We further evaluate the method by calculating quantitative metrics on an expanded dataset of 50 facades and compare the results to the proposed ControlNet and Photoshop baselines.

\paragraph{Sliced Wasserstein Distance (SWD).} First, we extract mid-level VGG16 feature maps \cite{simonyan2014very} for each image, L2-normalize these features, and compute the Sliced Wasserstein Distance \cite{bonneel2015sliced} between the original image and each retargeted output. Specifically, we set the number of projection directions $n_{\text{projections}}=500$ to balance computational efficiency with the stability of the distance estimation. A lower SWD indicates higher perceptual similarity, capturing both color and textural fidelity.

\paragraph{CLIP Cosine Distance (CCD).} We measure semantic similarity with CLIP’s image encoder~\cite{Radford2021LearningTV}, using the ViT-B/32 backbone and its default preprocessing (resize and center-crop to $224{\times}224$). For each image, we extract the global image embedding and L2-normalize it. The distance between a ground-truth image $x$ and a retargeted output $y$ is defined as the cosine distance
\[
d_{\text{CLIP}}(x,y)=1-\langle \hat{e}_x,\hat{e}_y\rangle,
\]
where $\hat{e}_x$ and $\hat{e}_y$ are the normalized CLIP embeddings. Lower values indicate higher similarity. Unlike SWD, which is sensitive to local color/texture statistics, the CLIP distance emphasizes high-level, semantic agreement (objects and scene layout) while remaining agnostic to low-level pixel differences.

Table~\ref{tab:quntitative} showcases the means and standard deviations of both metrics across the whole set of facades for each of the compared methods. Our approach  yields lower Sliced Wasserstein Distance values than the ControlNet and Photoshop baselines, indicating superior preservation of local appearance and architectural detail.
Our method strongly outperforms the ControlNet baseline when it comes to the CLIP Cosine Distance, showing clear superiority to another fully generative approach. The CCD value is slightly lower than our approach in the case of the Photoshop baseline. This is to be expected, since the baseline is created by directly copying the full objects from the original image which the CLIP metric highly rewards. Despite that our CCD score is close behind proving a good level of semantic agreement.
Overall, these results correspond to the user study findings and indicated measurable improvement in retaining facade identity.

\begin{table}[b]
\centering
\caption{Quantitative evaluation of performance of our method against the ControlNet Baseline and the Photoshop Baseline. The table showcases the mean values and standard deviations of both the Sliced Wasserstein Distance and CLIP Cosine Distance metrics for all the methods. Note that the CCD value for Photoshop is lowest, which is expectable as the baseline is created by directly copying the full objects from the original (cf. Section~\ref{sec:quantitative}).}
\label{tab:quntitative}
\small
\setlength{\tabcolsep}{6pt}
\renewcommand{\arraystretch}{1.2}
\begin{tabular}{@{}l|cc|cc@{}}
\hline
 & \multicolumn{2}{c|}{\textbf{SWD}\,$\downarrow$} &
   \multicolumn{2}{c}{\textbf{CCD}\,$\downarrow$} \\
 & mean & std. & mean & std. \\
\hline
ControlNet Baseline & 0.1844 & 0.0387 & 0.2146 & 0.0769 \\
Photoshop Baseline & 0.124 & 0.0256 & 0.0936 & 0.043 \\
\hline
\textbf{Our Method}       & 0.1185 & 0.0215 & 0.118 & 0.0531 \\
\hline
\end{tabular}
\end{table}

\subsection{Ablations}

\begin{figure}[t]
\centering
 \input{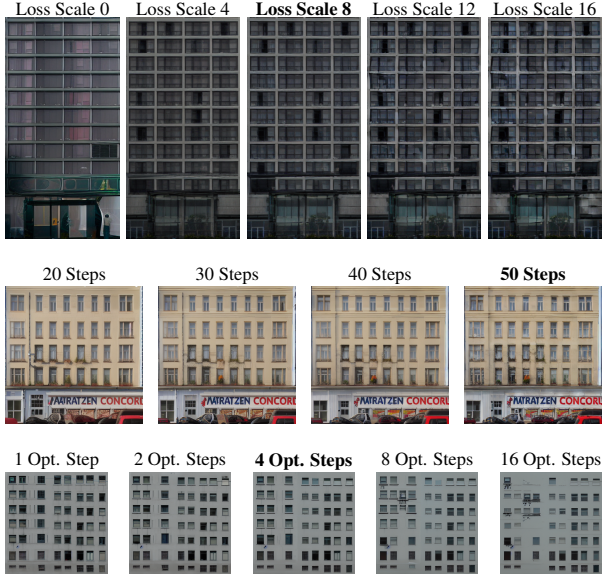}
\caption{Hyperparameter Ablations. We systematically vary three key hyperparameters: optimization step size (top row), diffusion guidance steps count (middle row), and optimization steps count (bottom row). Each row depicts the effect of altering one of the parameters. Bolded values indicate best-performing configurations.}
\label{fig:ablations_1}
\end{figure}

\paragraph{Hyperparameters.}
We perform various parameter ablations on the key parameters having the most influence over the quality of the method's results. The results, along with the optimal values, are highlighted in Figure~\ref{fig:ablations_1}.
\begin{itemize}

\item \textbf{Optimization Step Size.}  
Increasing the optimization step size encourages the model to retain more features of the original design by converging faster to the target activations. However, setting it too may lead to artifacts when overshooting the local minimum.

\item \textbf{Diffusion Guidance Steps.}  
We guide the process through all 50 denoising steps, unlike DiffusionHandles \citet{pandey2024diffusionhandles} (which stops at step 38). This extended guidance yields sharper and more refined results, likely because our method remodels the entire image space, as opposed to limiting itself to a single object.

\item \textbf{Optimization Steps Count.}  
Increasing the number of optimization steps at each denoising iteration while reducing their magnitude avoids artifacts and improves fidelity, albeit at higher computation time. Too many small steps can also cause the method to overfit to the original activations.

\end{itemize}

\paragraph{Module Ablations.}
We evaluate how the model performs when certain elements of the pipeline are disabled to highlight their impact on the performance of the whole method (Figure~\ref{fig:ablations_2}).

\begin{itemize}
\item \textbf{ControlNet and Activations Guidance.}  
We demonstrate the impact of omitting each guidance module from the pipeline. Without ControlNet, activations alone do a good job at transferring the core identity of the original facade design but cause hallucinations in certain areas of the resulting image.On the other hand, removing activations guidance, while keeping the ControlNet module, leads to blurry, uncertain outputs.

\item \textbf{Terminal Guidance.}  
Using the proposed hierarchical matching algorithm to pair every terminal symbol ensures comprehensive guidance across the whole facade. Disabling a specific terminal category guidance, like for walls or windows, can show how the core identity of the resulting design degrades when compared to the input; e.g., removing wall guidance makes the result lose the original brick texture (Figure~\ref{fig:ablations_2}). This highlights the importance of hierarchical matching, since correctly pairing empty regions like walls is possible thanks to the grammar based hierarchical representation of the structures.

\item \textbf{Canny ControlNet Finetuning.}  
We showcase how finetuning the Canny edges ControlNet model on curated dataset of facades influences the results of the whole pipeline. The results, while sometimes sharper (Figure~\ref{fig:ablations_2}), do not offer significant leaps in performance, making this step optional if resources are limited.
\end{itemize}

\begin{figure}[t]
\centering
\input{figures/figure_ablations_2}
\caption{Module Ablations. We perform an on/off analysis of our design components: ControlNet and activation guidance (top row), Wall terminals guidance (middle row) and ControlNet finetuning (bottom row). The top row illustrates results without using ControlNet or activation guidance, as well as full model results. The middle row highlights how the model performs with guidance disabled for specifically wall terminals. The bottom row illustrates the influence of using a ControlNet finetuned for facade images.}
\label{fig:ablations_2}
\end{figure}

\begin{figure}[b]
 \input{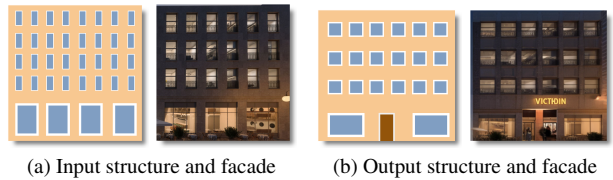}
        \caption{The showcase of a limitation of the method present when the user adds a brand new element to the output structure which was not present in the original. Since the doors never existed in the original image the algorithm has no reference so it improvises what should appear in the center area of the ground floor.}
        \label{fig:failure}
\end{figure}

\begin{figure}[t]
        \center
        \hfill
        \begin{subfigure}[t]{0.23\textwidth}
        \centering
        \includegraphics[width=0.48\textwidth]{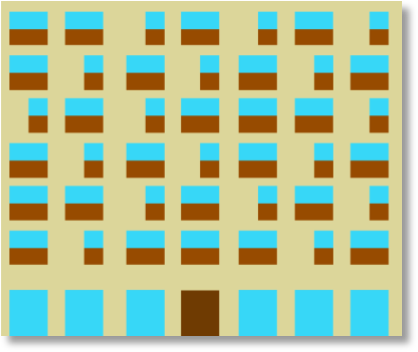}
        \includegraphics[width=0.48\textwidth]{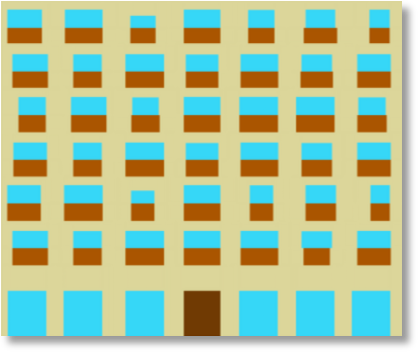}
        \caption{Limitation case}
        \label{fig:fail_proc_limitation}
        \end{subfigure}
        \hfill
        \begin{subfigure}[t]{0.23\textwidth}
        \centering
        \includegraphics[width=0.48\textwidth]{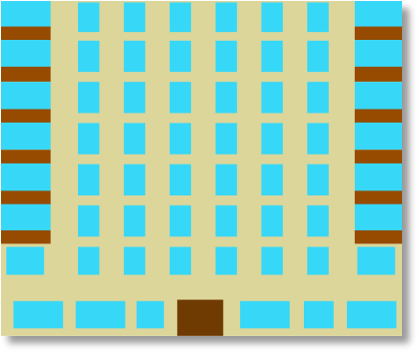}
        \includegraphics[width=0.48\textwidth]{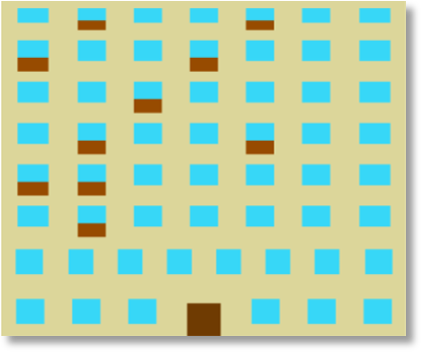}
        \caption{Failure case}
        \label{fig:fail_proc_failure}
        \end{subfigure}
        \hfill
        \caption{\begin{revision}Limitation and failure cases of the inverse procedural reconstruction module used in our pipeline (ground truth design and its inverse procedural reconstruction): (a) the underlying grammar is not expressive enough to reconstruct complex designs; (b) the model fails to recover the correct facade structure all together.\end{revision}}
        \label{fig:fail_proc}
\end{figure}

\begin{figure}[t]
    \center
    \hfill
    \begin{subfigure}[t]{0.32\textwidth}
    \centering
    \includegraphics[height=2cm]{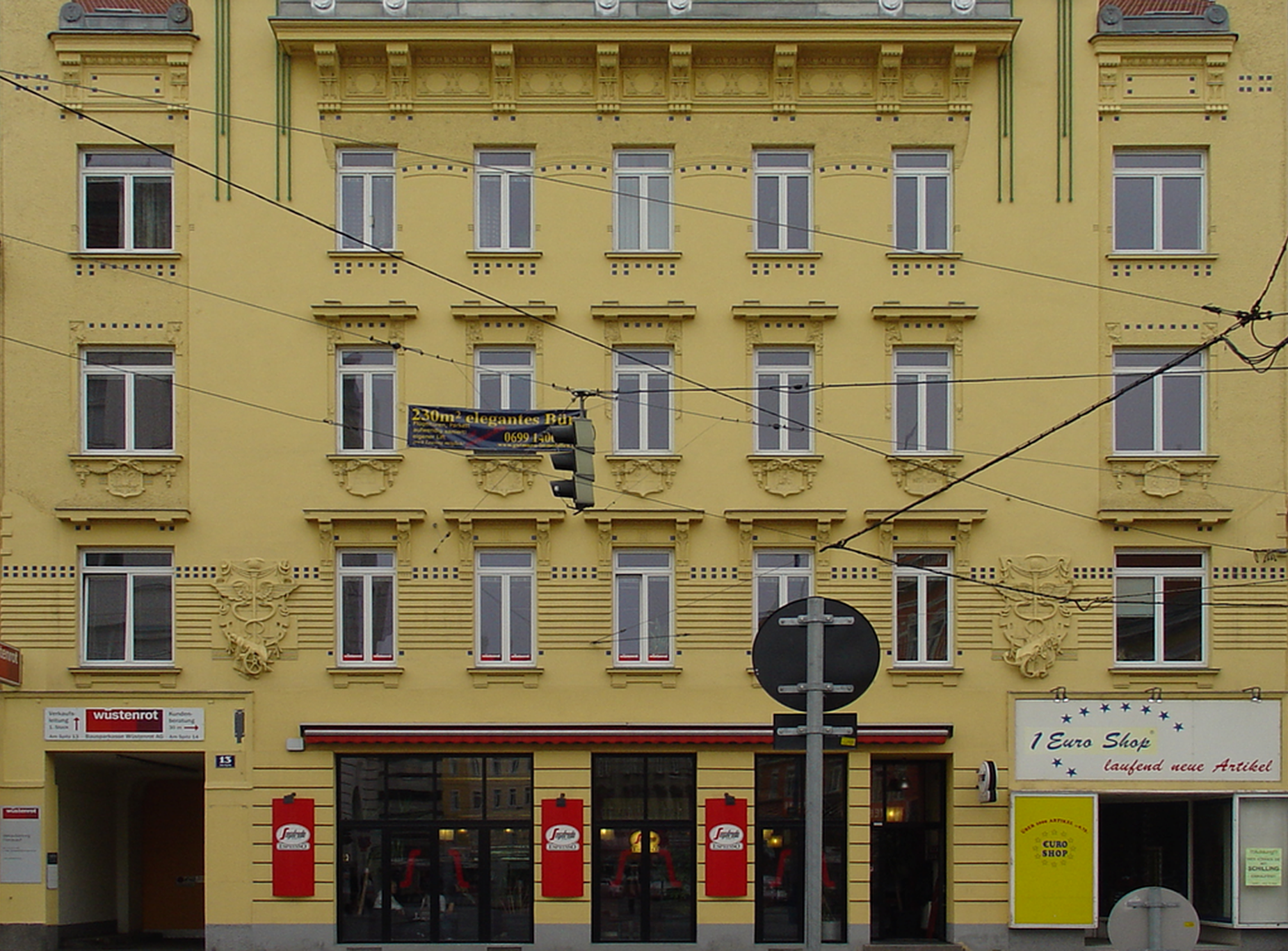}
    \includegraphics[height=2cm]{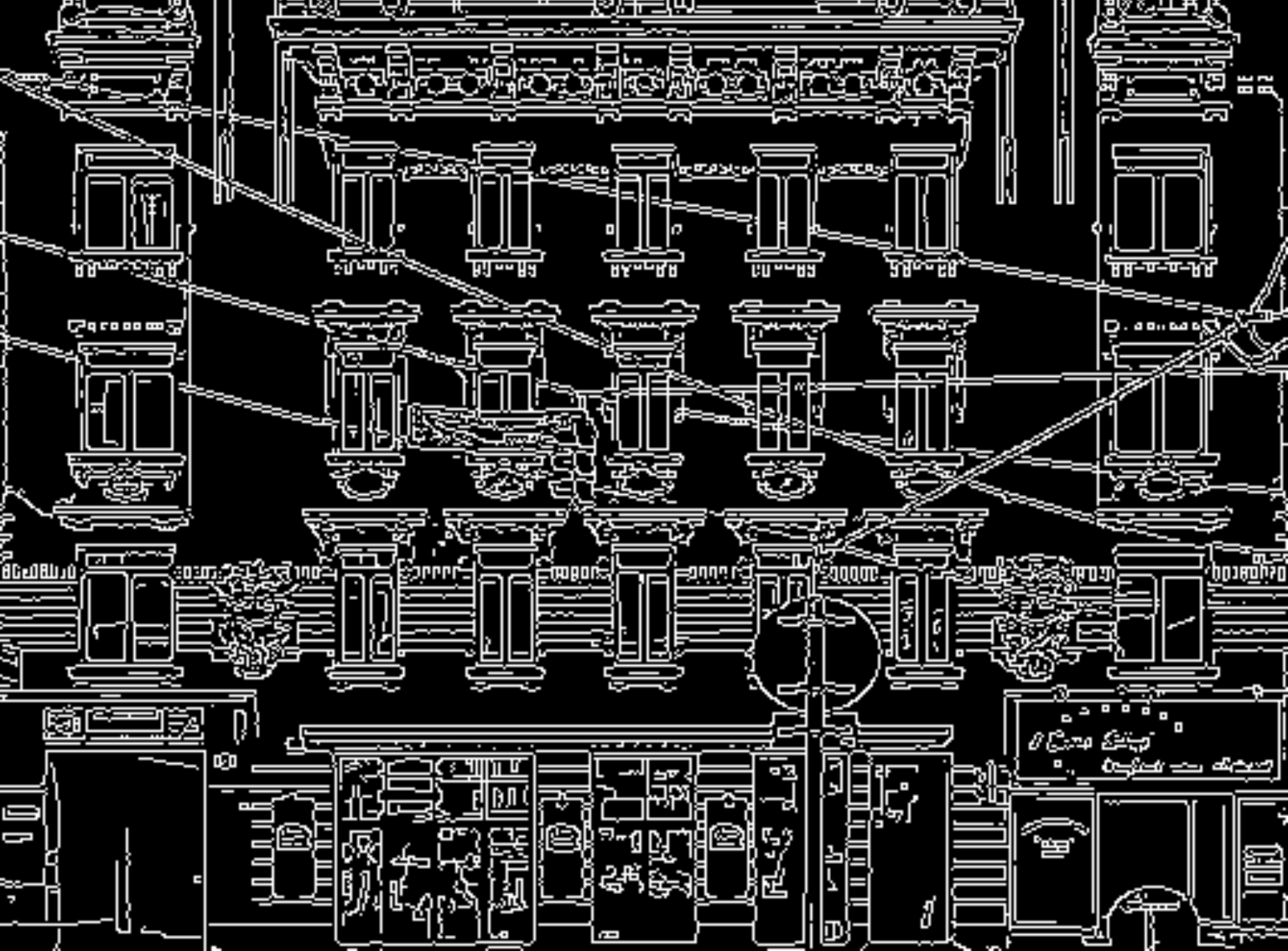}
    \caption{Input and generated Canny Edges}
    \label{fig:fail_in}
    \end{subfigure}
    \hfill
    \begin{subfigure}[t]{0.12\textwidth}
    \centering
    \includegraphics[height=2cm]{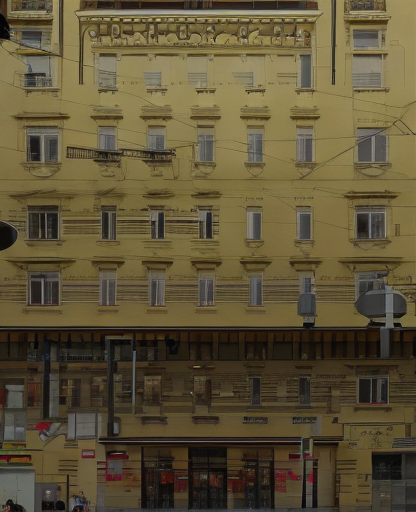}
    \caption{Output}
    \label{fig:fail_out}
    \end{subfigure}
    \hfill
    \caption{\begin{revision}Failure case example highlighting the influence of noise in the input (in the form of elements obstructing the facade design) on the final new variation generation result.\end{revision}}
    \label{fig:fail_noise}
\end{figure}

\subsection{Computational Efficiency \& Reproducibility}

\paragraph{Efficiency.}
All experiments were conducted on an NVIDIA RTX~3090 GPU. The one-time preprocessing—inverse procedural reconstruction (Fa\c{c}AID) and Null-text Inversion—takes on average 65\,s and 136\,s, respectively. Once computed, any number of edits can be performed on the same facade without re-running preprocessing. The cost of hierarchy matching averages around 19\,s. With preprocessing cached, a single guided-inference pass takes on average 56\,s to produce the final edited image. Peak memory usage is 14.9\,GB of VRAM, happening during the guided inference step.

\begin{revision}
\paragraph{Baseline Comparison.}
To provide context for these timings, we compare our computational cost against the baselines. The ControlNet baseline shares the one-time Null-text Inversion cost (136\,s), but its standard inference pass is faster (typically around 10 s) because it omits our iterative activation optimization. However, this speed comes at the significant cost of appearance preservation, as shown by our evaluations (e.g. Figure~\ref{fig:comparison_results}). Conversely, the Photoshop baseline is highly time-consuming, with manual collage assembly taking approximately 10 minutes per edit. Even when semi-automated using our hierarchical matching, the reliance on cloud-based generative fill (Adobe Firefly) introduces variable time overhead (on average 15\,s) and requires manual curation of variants, making it far less efficient for automated generation than our inference.
\end{revision}

\paragraph{Scalability.}
Preprocessing is a one-time cost per facade. Fa\c{c}AID scales with design complexity (more elements and deeper hierarchies take longer) and can fail on very complex layouts due to the sequence-length limits of its training. Null-text Inversion scales primarily with image resolution in a typical diffusion model fashion. The cost of hierarchical matching also grows with facade complexity, but in practice, Fa\c{c}AID will reach its complexity limit sooner than the cost of hierarchical matching becomes problematic. The guided-inference stage dominates the runtime when applying multiple edits to the same facade and grows mainly with the number of diffusion steps and inner optimization iterations, although increasing the number of these is generally unnecessary to obtain good results, making the performance of this step mostly dependent on the generated image resolution.

\paragraph{Reproducibility.}
All key parameters for inference, ControlNet fine-tuning and other implementation details are listed in Appendix~\ref{appx:implementation}. Additionally, Appendix~\ref{appx:metric} and Appendix~\ref{appx:matching} contain the pseudocode for key novel components. The source code will be available on the project's website.

\section{Limitations and Conclusions}\label{sec:conclustions}

\paragraph{Limitations.}
Although our method supports a wide range of facade edits, it struggles when new elements, which are not present in the original design, are introduced into the output hierarchical structure. For example, if the original facade lacks doors (Figure~\ref{fig:limitation_in}), adding door segments in the target structure (Figure~\ref{fig:limitation_out}) forces the hierarchical pairing to seek correspondences that simply do not exist. This results in missing valid control Canny edges and activations guidance in that region and can lead to visually inconsistent or incomplete areas in the resulting image.

A second limitation arises from the procedural language used by Fa\c{c}AID \cite{facaid2024}. The main failure mode occurs when the grammar on which the model was trained cannot express the structure of a given facade \begin{revision}(Figure~\ref{fig:fail_proc_limitation})\end{revision}. This is typical for facades with highly complex or unconventional element layouts. In such cases, the hierarchical structure produced by the model cannot faithfully approximate the real design because no valid derivation exists within the predefined grammar. Consequently, our method can only generate an approximate hierarchical representation constrained by the fixed grammar rules. This mismatch between the real and procedural designs leads to reduced hierarchical matching quality, as certain elements may be mislabeled or misaligned during the inverse procedural step.

Another type of failure occurs when the facade is theoretically representable by the grammar, but the inverse procedural reconstruction still fails to recover it \begin{revision}(Figure~\ref{fig:fail_proc_failure})\end{revision}. Since this step relies on a transformer-based model, occasional prediction errors are statistically unavoidable. In these cases, errors can accumulate and propagate through the pipeline, resulting in severely distorted or incoherent reconstructions.

\begin{revision}
Additionally, when the input facade image contains visual noise---objects that obstruct the building like trees, signs, cables, people, etc.---both the input Canny Edges map and the underlying network activations contain non-structural information (Figure~\ref{fig:fail_in}). When transferred to different parts of the image, that additional information becomes not cohesive and results in a noisy generation output (Figure~\ref{fig:fail_out}).
\end{revision}

Finally, our reliance on an early version of a diffusion model and the corresponding ControlNet backbone that supports Canny-edge conditioning---Stable Diffusion 1.5 and its corresponding SD~1.5 ControlNet---limits the achievable fidelity. Cutting-edge or specialized diffusion models without equivalent control interfaces could potentially yield higher realism. Thus, the generative quality of our outputs is bounded by the underlying model’s capacity. Adapting our pipeline to other diffusion backbones may enhance fine-detail rendering and style consistency. While the decision to stick with a simpler backbone was motivated mostly by available resources, very recent works like VideoHandles \cite{VideoHandles2025} show that our approach is fully adaptable to the new cutting edge wave of diffusion models---transformer based diffusion (DiT \cite{Peebles2022DiT}).

\paragraph{Conclusions.}
We presented a unified pipeline for realistic facade image editing that fuses shape grammar reconstructions with diffusion-based generation. By translating procedural edits into ControlNet guidance and activation alignment, our pipeline preserves the original facade’s core local appearance while enabling major structural modifications (e.g., multiplying floors, rearranging windows). The method, in contrast to most diffusion based editing approaches, does not merely focus on transforming a section of the image but provides a full image space reconstruction in a single pass, while still preserving appropriate fidelity. Our evaluations demonstrate that our method outperforms the presented baselines and showcase the capabilities of the proposed approach.

Looking ahead, we plan to explore advanced procedural grammars for more complex designs and alternative diffusion backbones that offer richer inference results. We believe that the synergy between symbolic procedural representations and data-driven synthesis holds promise for broader applications—from facade restoration to interactive architectural design—and demonstrates how neuro-symbolic integration can transform structured image editing.

\section*{Acknowledgments}
Generative AI tools were used for brainstorming and polishing the manuscript text. All scientific content and analysis were produced entirely by the authors.

\begin{figure*}[h]
    \centering
    \input{figures/figure_results_CGF.tex}
    \caption{Results of our method: Each row of five images begins with the original target image, followed by segmentation of variation 1 and its corresponding result, and then segmentation of variation 2 and its corresponding result. Best viewed in close-up in the electronic version.}
    \label{fig:qualitative_results}
\end{figure*}

\bibliographystyle{eg-alpha-doi} 
\bibliography{example_paper}       

\clearpage
\appendix

\section{Hierarchical Matching Metric}
\label{appx:metric}
We provide the pseudocode of the proposed metric used during hierarchical matching. Algorithm~\ref{alg:svd} describes the implementation of the SVD part of the metric, while Algorithm~\ref{alg:histogram_metric} showcases the histogram based metric. Finally, Algorithm~\ref{alg:full_metric} defines the full region similarity metric.

\begin{algorithm}[h!]
\caption{SVD Metric $D_{\text{SVD}}(A,B,\epsilon)$ for Regions $A$ and $B$}
\label{alg:svd}
\DontPrintSemicolon
\KwIn{$A\in\mathbb{R}^{M_A\times N_A}$, $B\in\mathbb{R}^{M_B\times N_B}$, threshold $\epsilon>0$}
\KwOut{$D_{\text{SVD}}(A,B,\epsilon)$}

\SetKwFunction{StructComplexity}{StructComplexity}
\SetKwProg{Fn}{Function}{:}{}

\Fn{\StructComplexity{$X,\epsilon$}}{
  $(U,S,V^\top)\gets \operatorname{SVD}(X)$ \tcp*{$S=\mathrm{diag}(\sigma_1\ge\dots\ge\sigma_r)$}
  $M\gets \text{rows}(X)$; $N\gets \text{cols}(X)$; $r\gets \min(M,N)$\;
  $\text{tailMSE}\gets \dfrac{1}{MN}\sum_{i=1}^{r}\sigma_i^2$\;
  $n\gets 0$\;
  \While{$n<r$ \textbf{and} $\text{tailMSE}\ge \epsilon$}{
    $\text{tailMSE}\gets \text{tailMSE} - \dfrac{\sigma_{n+1}^2}{MN}$\;
    $n \gets n+1$\;
  }
  \Return $C'_\epsilon(X)\gets n + \dfrac{\text{tailMSE}}{\epsilon}$\;
}

$C'_A \gets \StructComplexity(A,\epsilon)$\;
$C'_B \gets \StructComplexity(B,\epsilon)$\;
\Return $D_{\text{SVD}}(A,B,\epsilon) \gets \big|C'_A - C'_B\big|$\;

\end{algorithm}

\begin{algorithm}[h!]
\caption{Histogram Metric $D_H$ for Regions $A$ and $B$}
\label{alg:histogram_metric}
\DontPrintSemicolon
\KwIn{Grayscale regions $A$ and $B$ (arrays of intensities in $[0,255]$), number of bins $K \in \mathbb{N}$}
\KwOut{$D_H(A,B) \in [0,1]$}

\BlankLine
\textbf{1. Build histograms.}\;
\Indp
Compute (unnormalized) histograms $h_A \in \mathbb{R}^K$ and $h_B \in \mathbb{R}^K$ over $A$ and $B$ with the same bin edges.\;
\Indm

\BlankLine
\textbf{2. Normalize to probability masses.}\;
\Indp
$p_A \gets h_A / \sum_{i=1}^{K} h_A[i]$\;
$p_B \gets h_B / \sum_{i=1}^{K} h_B[i]$\;
\Indm

\BlankLine
\textbf{3. Bhattacharyya coefficient.}\;
\Indp
$\mathrm{BC} \gets \sum_{i=1}^{K} \sqrt{\,p_A[i]\;p_B[i]\,}$\;
\Indm

\BlankLine
\textbf{4. Hellinger distance (histogram metric).}\;
\Indp
\Return $D_H(A,B) \gets \sqrt{\,1 - \mathrm{BC}\,}$\;
\Indm

\end{algorithm}

\begin{algorithm}[h!]
\caption{Full Region Similarity Metric $D(A,B,\epsilon)$}
\label{alg:full_metric}
\DontPrintSemicolon
\KwIn{Regions $A,B$; SVD threshold $\epsilon>0$; histogram bins $K$; weights $\alpha,\beta\ge 0$}
\KwOut{$D(A,B)\in\mathbb{R}_{\ge 0}$}

\SetKwFunction{DSVD}{DSVD}
\SetKwFunction{DH}{DH}

$d_{\text{SVD}} \gets D_{\text{SVD}}(A,B,\epsilon)$\;
$d_{H} \gets D_{H}(A,B)$\;
\Return $D(A,B,\epsilon) \gets \alpha\cdot d_{\text{SVD}} + \beta\cdot d_{H}$\;

\end{algorithm}

\section{Hierarchical Matching Algorithm}
\label{appx:matching}
We provide the pseudocode of the hierarchical matching algortihm using the metric defined in Section~\ref{subsec:matching} and Appendix~\ref{appx:metric}.

\begin{algorithm}[h!]
\caption{Hierarchical Symbol Matching between $P_{\text{in}}$ and $P_{\text{out}}$}
\label{alg:hier_match}
\DontPrintSemicolon
\KwIn{Rooted symbol trees $P_{\text{in}},P_{\text{out}}$; region map $r(s)$; function $\operatorname{cat}(s)$; weights $\alpha,\beta>0$; SVD threshold $\epsilon>0$; Histogram bins $K \in \mathbb{N}$}
\KwOut{Set of pairs $\mathcal{M} \subseteq \text{nodes}(P_{\text{out}})\times \text{nodes}(P_{\text{in}})$}

\SetKwFunction{Compat}{Compatible}
\SetKwProg{Fn}{Function}{:}{}

\BlankLine
\Fn{\Compat{$s^{\text{out}}, s^{\text{in}}$}}{
  \Return $\big[\operatorname{cat}(s^{\text{out}})=\operatorname{cat}(s^{\text{in}})\big]$\;
}

\BlankLine
\textbf{Initialization.}\;
$\mathcal{M}\gets \emptyset$\;
mark all nodes in $P_{\text{in}}$ as unused\;
Order nodes of $P_{\text{out}}$ in breadth-first (root-to-leaves) order: $[s^{\text{out}}_1,\dots,s^{\text{out}}_T]$\;

\BlankLine
\For{$t \gets 1$ \KwTo $T$}{
  $s \gets s^{\text{out}}_t$\;
  $p_{\text{out}} \gets \text{parent}(s)$ \tcp*{$\bot$ for root}
  \eIf{$p_{\text{out}}=\bot$ \textbf{or} $(p_{\text{out}},p_{\text{in}})\notin \mathcal{M}$}{
    $\mathcal{C} \gets \{ u \in \text{nodes}(P_{\text{in}}) \mid \Compat(s,u)\ \wedge\ u \text{ unused} \}$
  }{
    $p_{\text{in}} \gets \text{match of } p_{\text{out}} \text{ in } \mathcal{M}$\;
    $\mathcal{C} \gets \{ u \in \text{nodes}(\text{subtree}(p_{\text{in}})) \mid \Compat(s,u)\ \wedge\ u \text{ unused} \}$
  }
  \If{$\mathcal{C}=\emptyset$}{\textbf{continue}}
  $u^\star \gets \arg\min_{u\in\mathcal{C}} D\!(r(u),\, r(s), \epsilon)$\;
  $\mathcal{M} \gets \mathcal{M}\cup\{(s,u^\star)\}$\;
  mark $u^\star$ as used\;
}
\Return $\mathcal{M}$\;

\end{algorithm}

\section{Implementation Details}
\label{appx:implementation}
This sections provides additional details needed for full method implementation. It mentions the values for parameters we have used to conduct the presented experiments as well as clarifies some workflow details.

\subsection{Metric Parameters}
\begin{itemize}
    \item For the SVD Metric we use a threshold $\epsilon = 5$.
    \item For the Histogram Metric we use bin count $K = 16$.
    \item For the full Similarity Metric we use coefficients $\alpha = \beta = 0.5$.
\end{itemize}

\subsection{ControlNet Fine-Tuning} We fine-tune a ControlNet model on 23k facade images from the LSAA dataset \citep{lsaa}. As a starting point for training the model we use the already trained all-purpose official Canny Edges conditioned ControlNet model based on Stable Diffusion 1.5.

\paragraph{Dataset Preprocessing.} In order to use the  LSAA facades for training we needed to compute the Canny Edges map for each of the images. To give the model more data and not overfit to one level of detail we generated three different Canny Edges maps for each facade in the dataset using 3 different pairs of tresholds: (25, 75); (50, 150); and (100, 250).

\paragraph{Fine-Tuning Hyperparameters.} We use the official Diffusers ControlNet Stable Diffusion 1.5 training script available in their repository in order to fine-tune the model. We mostly use the default parameters from the script. We changed the number of epochs to 20 and the batch size to 8 in order to accommodate our computational capabilities. We then trained the model using the preprocessed dataset of 69k datapoints in total and starting the training process from the all-purpose official Canny Edges ControlNet model.

\subsection{Optimization Convergence}
During the guided inference step we perform an optimization scheme to sway the activation values according to our hierarchical mapping. When performing this optimization we don't check any optimization convergence criteria. At each of the 50 diffusion steps we perform a constant number of 4 optimization steps which adjust the current network activation values.

\section{User Study Details}
\label{appx:userstudy}
\begin{figure*}[t]
\captionsetup[subfigure]{labelformat=empty}
\centering
    \begin{subfigure}[b]{0.44\textwidth}
        \centering
        \includegraphics[height=8cm]{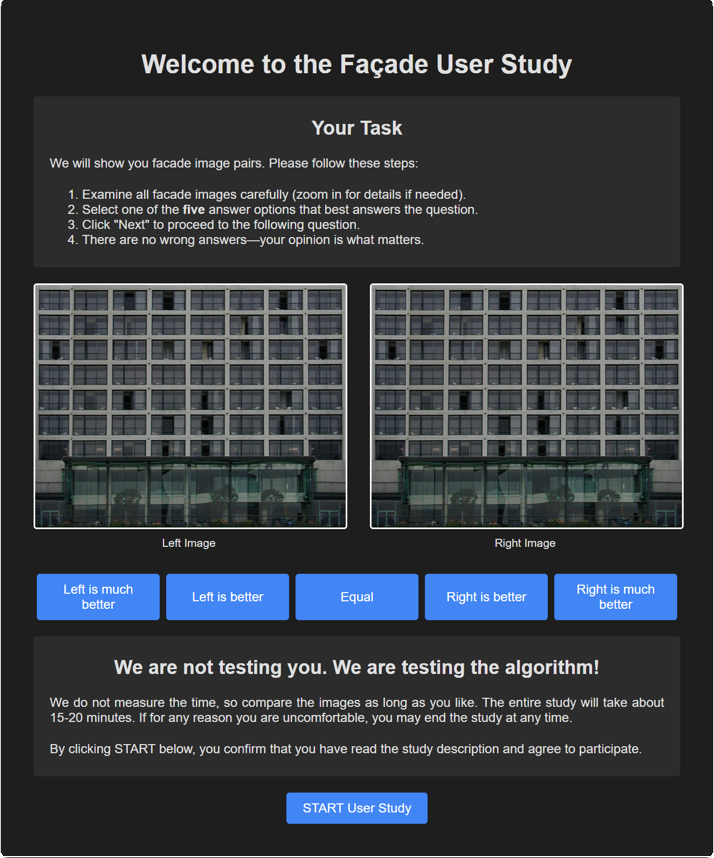} %
        \caption{(1) Landing Page}
        \label{fig:sub1}
    \end{subfigure}
    \hfill
    \begin{subfigure}[b]{0.44\textwidth}
    \centering
        \includegraphics[height=8cm]{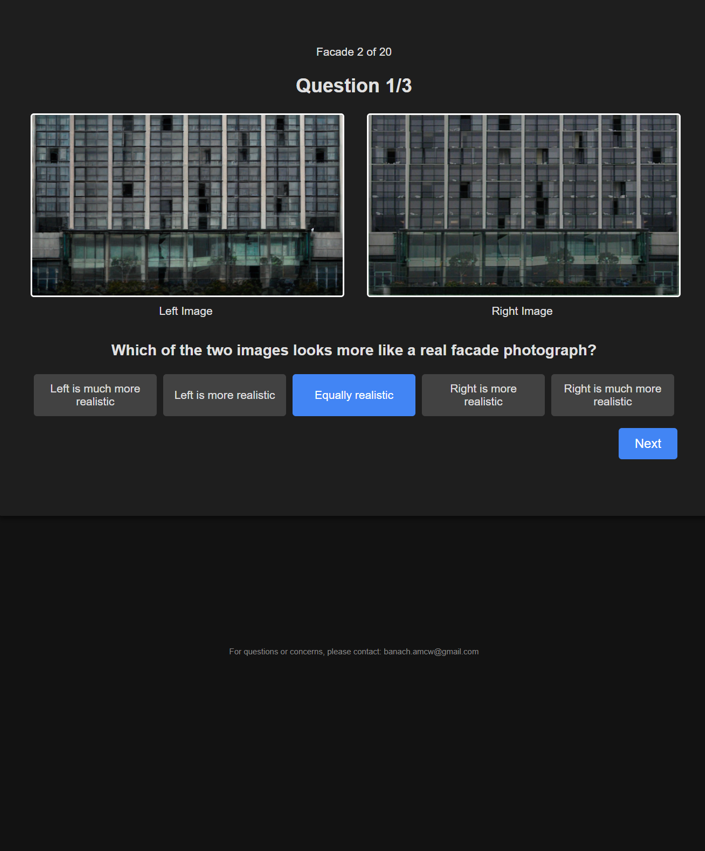} %
        \caption{(2) Realism}
        \label{fig:sub2}
    \end{subfigure}
    \hfill
    \begin{subfigure}[b]{0.44\textwidth}
    \centering
        \includegraphics[height=7.5cm]{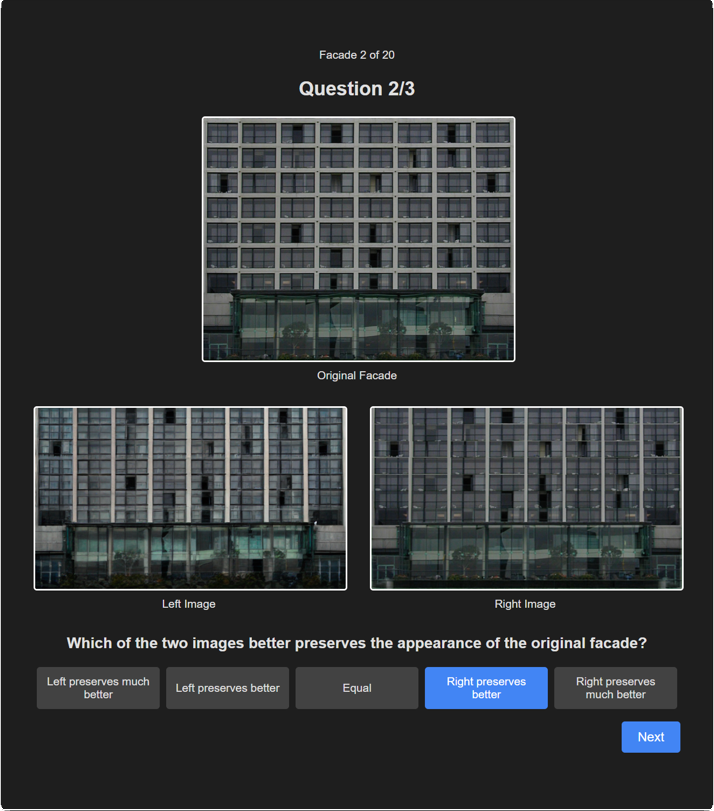} %
        \caption{(3) Appearance Preservation}
        \label{fig:sub3}
    \end{subfigure}
    \hfill
    \begin{subfigure}[b]{0.44\textwidth}
    \centering
        \includegraphics[height=7.5cm]{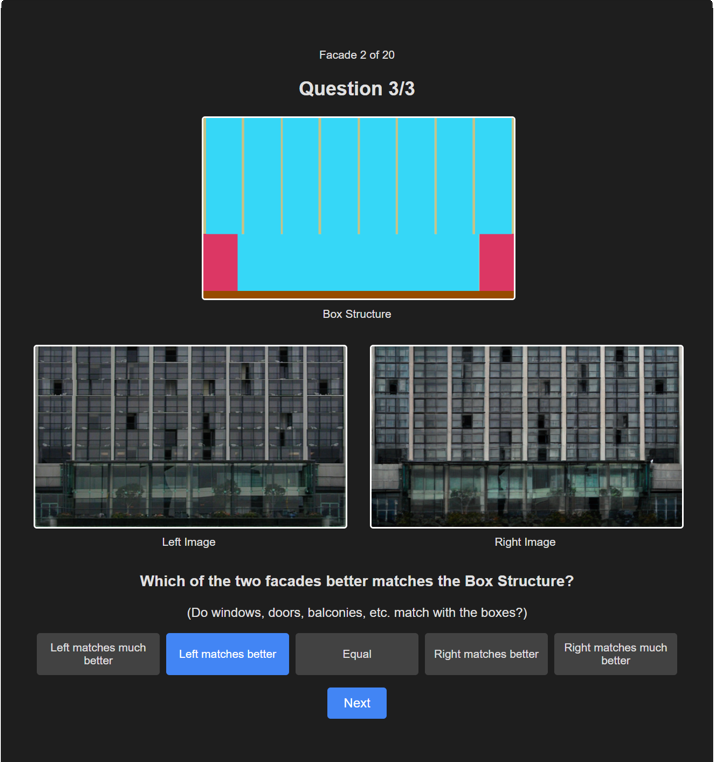} %
        \caption{(4) Edit Adherence}
        \label{fig:sub4}
    \end{subfigure}
    
    \caption{Showcase of how the user study looked like to the user: (1) landing introductory page; (2) realism question; (3) appearance preservation question; (4) edit adherence question.}
    \label{fig:app_user_study}
\end{figure*}

Each user was assigned a random subset of facade variations---10, 15 or 20, depending on the study group. For each selected variation the user was asked all three of the relevant questions: (1) realism question; (2) appearance preservation question; (3) edit adherence question. When answering any of the questions, the user could either express strong or mild preference to one of the results or deem them equal. The order in which the results were displayed (left or right position) was randomized for each question to keep the users from developing a preference towards any of the sides when answering multiple question in a row. Each question required the user to examine the result for at least 10 seconds before locking in the final answer. Figure~\ref{fig:app_user_study} presents how the user study looked like from the user's perspective.

\newpage

\end{document}